\documentclass[fleqn,usenatbib,useAMS]{mnras}

\usepackage{newtxtext,newtxmath}
\usepackage[T1]{fontenc}
\usepackage{xcolor, color}

\usepackage[utf8]{inputenc}
\usepackage{amsmath}
\usepackage[shortlabels]{enumitem}
\usepackage{graphicx}
\usepackage{color}
\usepackage{acro}
\usepackage{xcolor, color}
\usepackage{bm}

\newcommand{\rmd}{\textrm{d}}
\newcommand{\omgs}{\Omega_{\rm s}}
\newcommand{\omgc}{\Omega_{\rm c}}

\newcommand{\bs}[1]{\boldsymbol{#1}}
\newcommand{\tauc}{\tau_{\rm c}}
\newcommand{\taus}{\tau_{\rm s}}
\newcommand{\Ns}{N_{\rm s}}
\newcommand{\Nc}{N_{\rm c}}
\newcommand{\Is}{I_{\rm s}}
\newcommand{\Ic}{I_{\rm c}}

\newcommand{\expdiff}{e^{-(t-t')/\tau}}
\newcommand{\transition}{\bs{F}}
\newcommand{\processcovar}{\bs{Q}}
\newcommand{\measurecovar}{\bs{R}}
\newcommand{\emission}{\bs{B}}
\newcommand{\torques}{\bs{N}}
\newcommand{\bx}{\bs{x}}
\newcommand{\bxk}{\bx_{k}}
\newcommand{\hbxk}{\hat{\bx}_k}
\newcommand{\hbxkmo}{\hat{\bx}_{k-1}}

\newcommand{\by}{\bs{y}}
\newcommand{\byk}{\by_{k}}

\newcommand{\Nsimulations}{5000}

\title[Parameter estimation of a neutron star model]{Parameter estimation of a two-component neutron star model with spin wandering}
\author[P. M. Meyers et al.]{Patrick M. Meyers,$^{1,2}$\thanks{E-mail: pat.meyers@unimelb.edu.au}, Andrew Melatos$^{1,2}$, Nicholas J. O'Neill$^{1}$
\\
$^{1}$School of Physics, University of Melbourne, Parkville, VIC 3010, Australia\\
$^{2}$OzGrav, University of Melbourne, Parkville, VIC 3010, Australia
}
\date{Accepted XXX. Received YYY; in original form ZZZ}

\pubyear{2020}

\begin{document}
\label{firstpage}
\pagerange{\pageref{firstpage}--\pageref{lastpage}}

\maketitle
\begin{abstract}
It is an open challenge to estimate systematically the physical parameters of neutron star interiors from pulsar timing data while separating spin wandering intrinsic to the pulsar (achromatic timing noise) from measurement noise and chromatic timing noise (due to propagation effects).
In this paper we formulate the classic two-component, crust-superfluid model of neutron star interiors as a noise-driven, linear dynamical system
and use a state-space-based expectation-maximization method to estimate the system parameters using gravitational-wave 
and electromagnetic timing data. 
Monte Carlo simulations show that we can accurately estimate all six parameters of the two-component model provided that electromagnetic measurements of the crust angular velocity, and gravitational-wave measurements of the core angular velocity, are both available.
When only electromagnetic data are available we can recover
the overall relaxation time-scale, the ensemble-averaged spin-down rate, and the strength of the white-noise torque on the crust.
However, the estimates of the secular torques on the two components and white noise torque on the superfluid are biased significantly.
\end{abstract}


\section{Introduction}
Rapidly rotating neutron stars that simultaneously emit detectable electromagnetic and gravitational radiation are promising targets for multimessenger experiments on the physics of bulk nuclear matter. On the electromagnetic front, high-precision pulsar timing at radio, X-ray, and other wavelengths is a mature endeavor which yields a wealth of information about neutron star interiors \citep{lyne_graham-smith_2012}. On the gravitational front, long-baseline interferometers such as the Laser Interferometer Gravitational Wave Observatory (LIGO) \citep{TheLIGOScientific:2014jea}, Virgo \citep{TheVirgo:2014hva}, and Kamioka Gravitational Wave Detector \citep{PhysRevD.88.043007} seek to measure persistent, quasimonochromatic, continuous-wave signals from neutron stars with long-duration, matched-filter searches \citep{2013PrPNP..68....1R}. A special focus of multimessenger studies is the equation of state of condensed nuclear matter \citep{2016PhR...621..127L} and its associated thermodynamic phases \citep{Alford2001} and dynamics \citep{2017IJMPD..2630015G}. Multimessenger constraints on the equation of state have been inferred from the joint detection of the compact binary coalescence GW170817 and gamma-ray burst GRB 170817A by estimating the masses, radii, and tidal deformabilities of the merging neutron stars within a Bayesian framework \citep{2017PhRvL.119p1101A,2017ApJ...848L..12A,2018PhRvL.121p1101A,2019MNRAS.489L..91C}. Additional constraints come from recent mass and radius measurements of rotation-powered millisecond pulsars by the Neutron Star Interior Composition Explorer~\citep{10.1117/12.2231304}, inferred by modeling thermal X-ray pulse profiles and emission hot spots \citep{Riley_2019,Raaijmakers_2019,Miller_2019,Bogdanov_2019}.

Dense matter inside a neutron star is thought to consist of two principal components: (i) superfluid neutrons; and (ii) superconducting protons, which are locked magnetically to electrons and the rigid crust \citep{1969Natur.224..872B,Mendell1991a}. The two components are coupled by vortex-mediated mutual friction \citep{1969Natur.224..872B,Mendell1991} and entrainment \citep{1976JETP...42..164A,2006CQGra..23.5505A}.
\footnote{Generalizations to three or more components also exist \citep{2006CQGra..23.5505A,2011MNRAS.410..805G,10.1093/mnras/sty2864}.}
Evidence for the two-component model comes from theoretical nuclear physics [\citet{1999PhyU...42..737Y,2017JApA...38...43C,2016PhR...621..127L} and references therein] and from the time-scale and quasiexponential form of rotational glitch recoveries [\citet{alpar:1984yui,2010MNRAS.409.1253V,2015IJMPD..2430008H} and references therein]. Historically the parameters of the two-component model have been estimated observationally in two independent ways: (i) by fitting timing models to individual glitch recoveries \citep{1986ApJ...311..197A,2010MNRAS.405.1061S,2010MNRAS.409.1253V,2011MNRAS.414.1679E,2013PhRvL.110a1101C,2016MNRAS.460.1201H,2018ApJ...865...23G,2019NatAs...3.1143A}; and (ii) by studying long-term spin down, as affected by the relative inertia of the superfluid and crust, for example \citep{2011MNRAS.414.1679E,2012PhRvL.109x1103A,2013PhRvL.110a1101C,2014MNRAS.444.1318B,2017MNRAS.471.4827G}.

Parameter estimation with the two-component model is often confounded by spin wandering, which is covariant with glitch recoveries and other slow spin-down perturbations on time-scales of months or longer. In this paper, we define `spin wandering' to mean achromatic fluctuations in pulse times-of-arrival (TOAs) that are intrinsic to the pulsar --- specifically the rotation of its crust and magnetosphere --- in contradistinction to chromatic TOA fluctuations caused by propagation through the magnetosphere and interstellar medium~\citep{2013MNRAS.429.2161K,2013CQGra..30v4002C,2014ApJ...790L..22A,2016ApJ...818..166L,2016MNRAS.458.2161L,2017ApJ...834...35L,2020arXiv200810562D,2020MNRAS.tmp.3250G}. Spin wandering is thought to be partly responsible for timing noise, a stochastic red noise process seen in the power spectral density of pulsar timing residuals, after the best-fitting ephemeris has been subtracted \citep{1975ApJS...29..453G,1980ApJ...237..216C,1994ApJ...422..671A,Shannon2010,2012MNRAS.426.2507P,2019MNRAS.487.5854N,2019arXiv191005961G,2019MNRAS.489.3810P,2020MNRAS.tmp..578L,parthasarathy:2020wel}.
%\footnote{Fluctuations in magnetospheric structure, possibly unrelated to the spin, may also contribute \citep{Lyne2010,2019MNRAS.485.3230S}.}
When one estimates internal neutron star parameters observationally, spin wandering is treated normally as a ``nuisance'' -- that is, a random feature which survives after a deterministic timing model is fitted, and effectively limits the accuracy with which the underlying stellar parameters are measured.

In this paper we reverse the paradigm by taking direct advantage of spin wandering to estimate the parameters of the two-component neutron star model. Stochastic fluctuations in the observed rotation of the rigid crust excite the coupled dynamics of the two-component stellar interior just like any glitch, except that the stochastic excitation is ongoing, whereas a glitch is impulsive. Therefore spin wandering is a direct, driven outcome of the deterministic two-component dynamics. Much can be learned physically by tracking it, even though the statistics of the random driver are unknown a priori; for example it is possible to infer the two-component coupling time-scale \citep{2012MNRAS.426.2507P}. One advantage is that the volume of useful data expands; one is not limited to the immediate aftermath of a glitch to extract information about the model. Another advantage is that the tracking method generalizes naturally to multimessenger situations, where both electromagnetic data (which reflect the behavior of the crust) and gravitational wave data (which reflect the core as well) exist. However, the method is still viable when only electromagnetic data are available. To achieve the tracking, we employ state space identification and an expectation-maximization algorithm, which are exhaustively tested in the signal processing literature and widely used in practical engineering applications \citep{em_original_paper,Shumway1982,GIBSON20051667,durbin2012time}.

The paper is organized as follows. In Section \ref{sec:physical_model} we introduce and motivate an idealized physical model of a rotating neutron star, which has secular and stochastic torques applied to both the crust and core, which are coupled linearly as in the classic two-component model \citep{1969Natur.224..872B}. In Section \ref{sec:inferring_physical_parameters} we introduce a standard expectation-maximization algorithm to estimate the two-component model parameters from a uniform, discretely sampled time series of electromagnetic and/or gravitational wave data within a linear state space framework \citep{Shumway1982,Digalakis_1993_EM_algorithm}. In Section \ref{sec:simulated_data_results} we test the accuracy of the estimator under controlled conditions using Monte Carlo simulations to generate synthetic electromagnetic and gravitational wave data. The implications for future analyses with real, astronomical data are discussed briefly in Section \ref{sec:conclusion}.

\section{Two-component model}
\label{sec:physical_model}

\subsection{Equations of motion}
The two-component model of a rotating neutron star was introduced by \citet{1969Natur.224..872B} to describe the post-glitch recovery of the Vela pulsar. It divides the star into two parts: (i) a superfluid, which is widely believed to be an inviscid neutron condensate located in the inner crust or outer core; and (ii) a rigid, crystalline crust, which locks magnetically to the charged fluid species (e.g. electrons and superconducting protons) in the inner crust and outer core \citep{Mendell1991a,Mendell1991,2006CQGra..23.5505A,2011MNRAS.410..805G}. The components are assumed to rotate uniformly yet differentially, with angular velocities $\Omega_{\rm s}$ and $\Omega_{\rm c} \neq \Omega_{\rm s}$ respectively. They obey the coupled equations of motion
\begin{align}
 I_{\rm c} \frac{\rmd \Omega_{\rm c}}{\rmd t} 
 & =
 N_{\rm c} + \xi_{\rm c}(t) - 
 \frac{I_{\rm c}}{\tau_{\rm c}} (\Omega_{\rm c} - \Omega_{\rm s})~,
 \label{EQ:CRUST_EQ_OF_MOTION}
 \\
  I_{\rm s} \frac{\rmd\Omega_{\rm s}}{\rmd t} 
 & =
 N_{\rm s} + \xi_{\rm s}(t) - 
 \frac{I_{\rm s}}{\tau_{\rm s}} (\Omega_{\rm s} - \Omega_{\rm c})~,
 \label{EQ:SUPERFLUID_EQ_OF_MOTION}
\end{align}
where the subscripts `s' and `c' label the superfluid and crust respectively, $I_{\rm s}$ and $I_{\rm c}$ are effective moments of inertia, $N_{\rm s}$ and $N_{\rm c}$ are secular torques, $\xi_{\rm s}$ and $\xi_{\rm c}$ are stochastic torques, and $\tau_{\rm s}$ and $\tau_{\rm c}$ are coupling time-scales. The physical origin of the six torques on the right-hand sides of (\ref{EQ:CRUST_EQ_OF_MOTION}) and (\ref{EQ:SUPERFLUID_EQ_OF_MOTION}) are discussed further below in Sections \ref{ssec:deterministic_torques} and \ref{ssec:stochastic_torques}.

We emphasize at the outset that the model expressed by (\ref{EQ:CRUST_EQ_OF_MOTION}) and (\ref{EQ:SUPERFLUID_EQ_OF_MOTION}) is highly idealized. For example, the assumption that the components individually rotate uniformly breaks down for $\Omega_{\rm c} \neq \Omega_{\rm s}$; Ekman pumping drives meridional flows in both components \citep{Reisenegger:1993aab,Abney:1996aaaa,Peralta2005,2010MNRAS.409.1253V,van_Eysden_2013} and even Kolmogorov-like turbulence \citep{GREENSTEIN_1970,Melatos_2007}. In the absence of uniform rotation, the quantities $I_{\rm s}$, $\Omega_{\rm s}$, $I_{\rm c}$, and $\Omega_{\rm c}$ represent body-averaged approximations to the realistic behavior \citep{2010MNRAS.405.1061S,haskell:2012rds}; see also \citet{Antonelli:2017oxd} for an axially symmetric generalization. Moreover the superfluid is pinned to nuclear lattice sites \citep{Link:1991dlp,Avogadro:2008poi,Seveso:2016wqd} and quantized magnetic fluxoids \citep{Bhattacharya:1991lsds,ruderman:1998ids,sidery:2009kds,2017MNRAS.471.4827G,drummond:2017wfg} in the inner crust, leading to non-linear stick-slip dynamics in $\Omega_{\rm s}$ which do not emerge explicitly from (\ref{EQ:SUPERFLUID_EQ_OF_MOTION}). Some aspects of the stick-slip dynamics are captured phenomenologically by hydrodynamic models \citep{haskell:2013sdd,khomenko:2018sdf}, but others are not, e.g.\ the long-term statistics of vortex unpinning events \citep{Warszawski:2011ldo,melatos:2013pwe,fulgenzi:2017lpw,Melatos:2018lsd} which drive the behavior of $\xi_{\rm s}$. Finally equations (\ref{EQ:CRUST_EQ_OF_MOTION}) and (\ref{EQ:SUPERFLUID_EQ_OF_MOTION}) omit magnetohydrodynamic forces \citep{2011MNRAS.410..805G}, which are important dynamically with respect to the small lag $|\Omega_{\rm s} - \Omega_{\rm c}| \ll \Omega_{\rm s,c}$ and behave subtly, when the geometry of the internal magnetic field is complicated \citep{Easson:1979edf,melatos:2012ped,glampedakis:2015yhf,drummond:2018esf,2020MNRAS.494.3095A}.

\subsection{Deterministic torques}
\label{ssec:deterministic_torques}
The physical interpretation and sign of the secular torques $N_{\rm s}$ and $N_{\rm c}$ depend on whether the neutron star is isolated or accreting. Consider first the crust. In all scenarios it corotates with the large-scale stellar magnetic field and spins down via magnetic dipole braking \citep{goldreich:1969wiu}. In addition it may possess a thermally or magnetically induced mass quadrupole moment, which exerts a gravitational radiation reaction torque \citep{ushomirsky:2000bvc,melatos:2005yup}, whose magnitude depends on the crust's composition and hence whether it is accreted or not \citep{haskell:2006oiu,priymak:2011try}. The above effects imply $N_{\rm c} < 0$. However, if the star accretes while it is being observed, the crust experiences a hydromagnetic accretion torque as well, whose sign depends on the detailed geometry of the magnetosphere-accretion-disk interaction \citep{Ghosh1979,1997ApJS..113..367B,romanova2004khj}. When the accretion torque is added, the net torque satisfies $N_{\rm c} < 0$ or $N_{\rm c} > 0$, depending on the effective averaging time-scale in the two-component model~\citep{1997ApJS..113..367B} and whether or not it brackets multiple cycles of quiescence and activity \citep{melatos:2016aaf}.

Next consider the neutron superfluid. It is decoupled electromagnetically from the crust so it does not experience the magnetic dipole braking torque directly. However it may possess a time-varying mass or current quadrupole moment, which exerts a gravitational radiation reaction torque with $N_{\rm s} < 0$, e.g.\ because the superfluid phase is inhomogeneous \citep{alpar:1996aew,sedrakian:2002vbf,2010MNRAS.402.2503J,2019AIPC.2127b0007H} or turbulent \citep{melatos:2010bvy,Melatos2014}. If the superfluid is pinned strongly, its angular velocity evolves in a step-wise fashion, with ${\rm d}\Omega_{\rm s}/{\rm d}t = 0$ instantaneously except at stick-slip events like vortex avalanches \citep{Warszawski:2011ldo}. However, if the averaging time-scale in the hydrodynamic two-component model is longer than the typical inter-avalanche waiting time, we have $\rmd \Omega_{\rm s}/\rmd t < 0$ and $N_{\rm s} < 0$ in (\ref{EQ:CRUST_EQ_OF_MOTION}) and (\ref{EQ:SUPERFLUID_EQ_OF_MOTION}), e.g. due to vortex creep \citep{alpar:1984yui,alpar:1984hrw,sedrakian:2002vbf,sidery:2009kds,Gugercinoglu:2014als}. Accretion does not modify $N_{\rm s}$ to a good approximation.

The rightmost terms in (\ref{EQ:CRUST_EQ_OF_MOTION}) and (\ref{EQ:SUPERFLUID_EQ_OF_MOTION}) couple the crust and superfluid. They act to reduce the lag $| \Omega_{\rm s} - \Omega_{\rm c} |$ and form an action-reaction pair (Newton's Third Law) in the special case $I_{\rm s} / \tau_{\rm s} = I_{\rm c} / \tau_{\rm c}$.
\footnote{
In general the star contains additional components, which store and release angular momentum \citep{alpar:1996aew}.
}
Physically the coupling arises from vortex-mediated mutual friction \citep{1969Natur.224..673B,Mendell1991} and entrainment \citep{1976JETP...42..164A,2006CQGra..23.5505A}. The restoring torque is linear in $| \Omega_{\rm s} - \Omega_{\rm c} |$ to a good approximation in the regime $|\Omega_{\rm s} - \Omega_{\rm c}| \ll \Omega_{\rm s,c}$, but other functional forms are possible too, e.g. $\propto | \Omega_{\rm s} - \Omega_{\rm c} |^3$ if the superfluid is turbulent \citep{GORTER1949285,Peralta2005}. The relaxation time-scales $\tau_{\rm s}$ and $\tau_{\rm c}$ typically run from days to weeks in theory \citep{Mendell1991} and are consistent with observed glitch recovery time-scales \citep{2010MNRAS.409.1253V} and the autocorrelation time-scale of spin wandering \citep{2012MNRAS.426.2507P}.

In this paper, $N_{\rm s}$ and $N_{\rm c}$ are treated as constant for the sake of simplicity and also because joint multimessenger observations are typically short ($\sim 1\,{\rm yr}$) compared to the secular spin-down or spin-up time-scale. In general one has $N_{\rm c} \propto \Omega_{\rm c}^3$ for magnetic dipole braking and $N_{\rm c,s} \propto \Omega_{\rm c,s}^5$ for gravitational radiation reaction among other effects. Solutions for the general case can be developed, if the data warrant \citep{2017MNRAS.471.4827G}.

\subsection{Stochastic torques}
\label{ssec:stochastic_torques}
Stochastic spin wandering is observed in isolated \citep{1975ApJS...29..453G,1980ApJ...237..216C,1994ApJ...422..671A,Shannon2010,2012MNRAS.426.2507P,2019MNRAS.487.5854N,2019arXiv191005961G,2019MNRAS.489.3810P} and accreting \citep{Baykal:1991uec,Baykal:1993aaa,deKool:1993wvd,baykal:1997lvw,1997ApJS..113..367B} neutron stars. In general it is driven by stochastic torques $\xi_{\rm s}$ and $\xi_{\rm c}$ acting on the superfluid and crust respectively in the two-component model (\ref{EQ:CRUST_EQ_OF_MOTION}) and (\ref{EQ:SUPERFLUID_EQ_OF_MOTION}), although only the response of the crust has been observed directly in the electromagnetic data available to date.

In this paper, for the sake of simplicity, we treat $\xi_{\rm s}(t)$ and $\xi_{\rm c}(t)$ as memoryless, white noise processes, whose ensemble statistics obey
\begin{align}
 \langle \xi_{\rm s,c}(t) \rangle
 & = 
 0~,
 \label{eq:white_noise_mean}
 \\
 \langle \xi_{\rm s,c}(t) \xi_{\rm s,c}(t') \rangle
 & = 
 \sigma_{\rm s,c}^2 \delta(t-t')~,
 \label{eq:white_noise_variance}
\end{align}
where $\langle \dots \rangle$ denotes the ensemble average, and $\sigma_{\rm s}$ and $\sigma_{\rm c}$ are noise amplitudes.
\footnote{
Equivalently one can replace (\ref{eq:white_noise_mean}) and (\ref{eq:white_noise_variance}) with a shot-noise model \citep{Baykal:1991uec}.
}
The quantities $\sigma_{\rm s}$ and $\sigma_{\rm c}$ can be inferred from pulse timing data, e.g.\ as described in Section \ref{sec:inferring_physical_parameters}, or they can be measured independently, e.g.\ from a periodogram analysis of the X-ray flux as in Scorpius X$-$1 \citep{Mukherjee:2018rim}. To gain a rough, intuitive sense of how (\ref{eq:white_noise_mean}) and (\ref{eq:white_noise_variance}) relate to observations, consider the special case $N_{\rm c}=0$ and $\tau_{\rm c} \rightarrow \infty$, which implies $\Omega_{\rm c}(t) = \int_0^t \rmd t' \, \xi_{\rm c}(t')$. Let $\delta\phi_{\rm c}(t) \propto \int_0^t \rmd t'' \, \Omega_{\rm c}(t'')$ be the residual rotational phase of the crust after subtracting the secular timing model inferred electromagnetically. Then the power spectral density of the phase residuals is computed from the autocorrelation function of $\delta\phi_{\rm c}(t)$ according to 
\begin{equation}
 \Phi(f)
 =
 \int_{-\infty}^\infty \rmd \tau \, 
 e^{2\pi i f \tau}  
 \langle \delta\phi_{\rm c}(t) \delta\phi_{\rm c}(t+\tau) \rangle
\end{equation}
and satisfies
\begin{equation}
 \Phi(f) \propto f^{-4}~.
 \label{eq:timing_noise_power_law}
\end{equation}
The power law (\ref{eq:timing_noise_power_law}) is a respectable fit to several accretion-powered X-ray pulsars, e.g.\ Her X$-$1, Cen X$-$3, A0535$+$26, and Vela X$-$1 \citep{Baykal:1993aaa}. It is also a respectable fit to many isolated radio pulsars \citep{Melatos2014,lasky:2015pfd,2019MNRAS.487.5854N,2019arXiv191005961G,2020MNRAS.tmp..578L,parthasarathy:2020wel}. However we emphasize that (\ref{eq:white_noise_mean}) and (\ref{eq:white_noise_variance}) are nothing more than phenomenological ans\"{a}tze introduced here to illustrate the potential of the data analysis method in the paper; they are not universal. Many neutron stars have $\Phi(f)$ shallower or steeper than $f^{-4}$ in (\ref{eq:timing_noise_power_law}); see \citet{Baykal:1993aaa}, \citet{Mukherjee:2018rim}, and the analysis of frequency and phase noise by \citet{1980ApJ...237..216C} and \cite{1980ApJ...239..640C}. Moreover $\Phi(f)$ may display non-power-law features, e.g. it may turn over at a characteristic frequency, something predicted in~\cite{Melatos2014} and searched-for explicitly by~\cite{2019arXiv191005961G}. Equations (\ref{EQ:SUPERFLUID_EQ_OF_MOTION})--(\ref{eq:white_noise_variance}) must be generalized to embrace these complications, if the data warrant.
\footnote{
\label{foot:freq_noise_footnote}
For example, frequency noise can be included by adding an equation of the form $d\phi_{\rm c}/dt = \Omega_{\rm c} + \zeta(t)$, where $\zeta(t)$ is a white-noise Langevin term obeying statistics like (\ref{eq:white_noise_mean}) and (\ref{eq:white_noise_variance}).
}

The specific physical interpretation of $\xi_{\rm s}$ and $\xi_{\rm c}$ depends on the astrophysical context. The pinned superfluid evolves step-wise through a sequence of metastable rotational states, with each step triggered by a vortex avalanche \citep{Warszawski:2011ldo,drummond:2018esf}, as discussed in Section \ref{ssec:deterministic_torques}. Although these stick-slip dynamics are driven ultimately by the crust spinning down or up,
\footnote{
Spin up (down) drives vortex avalanches radially inwards (outwards).
}
the vortex array stores stress in a complicated, time-dependent, many-vortex state and then releases it spasmodically and unpredictably. For example, the stress released in an avalanche may be less than, equal to, or greater than the stress accumulated since the last avalanche. The resulting torque $\xi_{\rm s}(t)$ is effectively random, and the crust feels a proportional random back reaction, $\xi_{\rm c}(t) = -\xi_{\rm s}(t)$. In addition shear-driven turbulence and meridional circulation generate random fluctuations in $\Omega_{\rm s}$, and fluctuations in the geometry of the magnetosphere contribute independently to $\xi_{\rm c}(t)$, triggered by vortex creep \citep{cheng:1987ekv}, plasma instabilities \citep{cheng:1987uid}, or global magnetic switching \citep{lyne:2010pds,stairs:2019jfs}. Magnetospheric fluctuations can also perturb the magnetic-field-aligned radio beam in stellar latitude and longitude, introducing phase noise independent of the underlying rotation and hence $\Omega_{\rm c}$. [The latter effect is omitted from (\ref{EQ:CRUST_EQ_OF_MOTION}) and (\ref{EQ:SUPERFLUID_EQ_OF_MOTION}) for the sake of simplicity in this paper; see footnote \ref{foot:freq_noise_footnote}.] Finally, if the star is accreting, fluctuations in the accretion torque dominate $\xi_{\rm c}(t)$. They are caused by hydromagnetic instabilities in the accretion-disk-magnetosphere interaction for example \citep{romanova2004khj,romanova:2008ieu,dangelo:2010esf}. The accretion-driven ensemble statistics of $\xi_{\rm c}(t)$ can be related approximately to X-ray flux variability via a simple mass-transfer model \citep{Mukherjee:2018rim}. One application of the parameter estimation technique presented in Section \ref{sec:inferring_physical_parameters} is to shed light on the relative contributions of the physical processes above (and others) to $\xi_{\rm s}$ and $\xi_{\rm c}$ by tracking spin wandering directly, not just measuring its ensemble statistics, and doing so self-consistently within the framework of (\ref{EQ:SUPERFLUID_EQ_OF_MOTION})--(\ref{eq:white_noise_variance}) \citep{Baykal:1993aaa}, instead of analysing $\Omega_{\rm c}(t)$ in isolation.

\subsection{Free modes}
\label{ssec:free_modes}
Equations (\ref{EQ:CRUST_EQ_OF_MOTION}) and (\ref{EQ:SUPERFLUID_EQ_OF_MOTION}) have an analytic solution, which is presented in Appendix~\ref{app:sec:analytic_solution}. The homogeneous system with $N_{\rm c,s}=\xi_{\rm c,s}=0$ has two modes that satisfy
\begin{align}
\tauc\omgc + \taus\omgs &= \textrm{ constant}
\label{eq:conservation_of_momentum}
\end{align}
and
\begin{align}
\frac{\rm d}{\textrm{d}t}\left(\omgs - \omgc\right) &= -\frac{\tauc + \taus}{\tauc\taus}\left(\omgs- \omgc\right).
\label{eq:exponential_damping}
\end{align}
Equation~(\ref{eq:conservation_of_momentum}) describes conservation of total angular impulse of the superfluid and crust. Equation~(\ref{eq:exponential_damping}) says that any crust-superfluid lag is exponentially damped on the characteristic time-scale
\begin{align}
\label{eq:tau}
\tau = \frac{\taus\tauc}{\taus+\tauc}.
\end{align}
Quasi-exponential recoveries are observed during the aftermath of many radio pulsar glitches~\citep{2010MNRAS.409.1253V}. The homogeneous solution has $\omgs \rightarrow \omgc$ as $t \rightarrow \infty$. The particular solution, shown in Appendix~\ref{app:sec:analytic_solution}~\citep{Baykal:1993aaa,2017MNRAS.471.4827G}, includes $N_{\rm c,s}\neq 0$ and integration over the history of $\xi_{\rm c,s}$. In this case, the lag may not tend to zero; indeed we have 
\begin{align}
\omgs(t) - \omgc(t) = \tau \left(\frac{\Ns}{\Is} - \frac{\Nc}{\Ic}\right) + \textrm{ stochastic terms}\label{eq:simple_model:lag_limit_long_time}
\end{align}
as $t\rightarrow \infty.$

It is also straightforward to see in the particular solution that as $t\rightarrow\infty$ the derivative of the ensemble-averaged angular velocity of the superfluid and the crust are equal and converge to
\begin{align}
\langle \dot\Omega_{\rm c}\rangle &= \langle\dot\Omega_{\rm s}\rangle \\
&= \frac{1}{\tauc + \taus}\left(\frac{\tauc \Nc}{\Ic} + \frac{\taus \Ns}{\Is}\right).
\label{eq:deterministic_spindown}
\end{align}

\section{Inferring Physical Parameters from Observations}
\label{sec:inferring_physical_parameters}
In this section we present a new method to simultaneously infer $\tauc$, $\taus$, $\Nc$, $\Ns$, $\sigma_{\rm c}$ and $\sigma_{\rm s}$. We do this using electromagnetic and gravitational-wave measurements to track the stochastic and secular components of $\omgc$ and $\omgs$ separately. This is in contrast to typical methods which fit the long-term secular behavior of the pulse arrival times and use the residuals to characterize the stochastic behavior, i.e. the timing noise.

Radio or X-ray observations are assumed to probe $\omgc$ because we expect electromagnetic emission to come from the crust itself or the magnetosphere. 
Gravitational-wave measurements are assumed to probe $\omgs$, under the assumption that the gravitational-wave emitting quadrupole moment is dominated by the superfluid.  These assumptions can be relaxed, and it is straightforward to accommodate alternative scenarios, e.g. where gravitational waves are emitted by a crust-related quadrupole.
Typical continuous gravitational-wave searches track wandering 
in the frequency of gravitational-wave emission~\citep{2016PhRvD..93l3009S,2017PhRvD..96j2006S,2018PhRvD..97d3013S}, which is usually a simple multiple of the rotation frequency. Pulsar timing studies return TOAs, 
which are used to reconstruct a phase model, which can be differentiated in turn to give $\omgc$. 

In Section~\ref{ssec:temporal_discretization} we present a discrete-time version of the model in (\ref{EQ:CRUST_EQ_OF_MOTION}) and (\ref{EQ:SUPERFLUID_EQ_OF_MOTION}), which maps onto the discretely sampled observables $\omgc(t)$ and $\omgs(t)$. In Section~\ref{ssec:kalman_filter} we discuss how observations of $\omgc(t)$ and $\omgs(t)$ can be used to estimate the smoothed state evolution of the discrete-time model using the Kalman filter, a classic tool from electrical engineering. In Section~\ref{ssec:infer:em_algorithm}, we present the expectation-maximization algorithm, which can be used to find the maximum-likelihood estimate of the model parameters, given the smoothed state evolution returned by the Kalman filter. Explicit details of the implementation of the expectation-maximization algorithm are left to Appendix~\ref{sec:em_overview_for_our_case} and Appendix~\ref{app:em_implementation}.

We emphasize again that the expectation-maximization algorithm treats the secular and stochastic components of the data as equally physical and fits both of them simultaneously. It does not treat the stochastic component as uninformative "noise" to be smoothed out. The measured stochastic response contains valuable information about deterministic physical processes, e.g. the white-noise drivers $\xi_{\rm c}(t)$ and $\xi_{\rm s}(t)$ are filtered by the deterministic coupling terms on the right-hand sides of (\ref{EQ:CRUST_EQ_OF_MOTION}) and (\ref{EQ:SUPERFLUID_EQ_OF_MOTION}).

\subsection{Temporal discretization}
\label{ssec:temporal_discretization}
Let subscript $k$ denote evaluation at time $t_k$, viz. $\Omega_{\textrm{c}, k} = \omgc(t_k)$.  Then (\ref{EQ:CRUST_EQ_OF_MOTION}) and (\ref{EQ:SUPERFLUID_EQ_OF_MOTION}) take the discrete form
\begin{align}
    \Omega_{\textrm c, k+1} &= \frac{\Nc}{\Ic}\Delta t +\Delta W_{\textrm c,k} + \left(1 - \frac{\Delta t}{\tauc} \right)\Omega_{\textrm c,k} + \frac{\Delta t}{\tauc}\Omega_{\textrm s,k}\label{eq:discrete:odec}\\
    \Omega_{\textrm s,k+1} &= \frac{\Ns}{\Is} \Delta t + \Delta W_{\textrm s,k} + \left(1 - \frac{\Delta t}{\taus}\right)\Omega_{\textrm s,k} + \frac{\Delta t}{\taus}\Omega_{\textrm c,k},\label{eq:discrete:odes}
\end{align}
to linear order in $\Delta t=t_{k+1} - t_k$, where 
\begin{align}
\Delta W_{\textrm c,k} = W_{\textrm c,k+1} - W_{\textrm c,k} 
\end{align}
represents a Wiener process
and $\Delta W_{\textrm s, k}$ likewise. In making this discrete time approximation we implicitly assume $\Delta t \ll \tauc,\,\taus,\Ic\omgc / \Nc$, and $\Is\omgs / \Ns$.

Equations (\ref{eq:discrete:odec}) and (\ref{eq:discrete:odes}) are of the linear form
\begin{align}
\label{eq:inference:state_transition_equation}
    \bs{x}_{k+1} &= \bs{F}\bs{x}_k + \torques + \Delta\bs{W}_k
\end{align}
with
\begin{align}
    \label{eq:discrete:xk_matrix}\bxk &= \begin{pmatrix}\Omega_{\textrm c,k}\\\Omega_{\textrm s,k}\end{pmatrix},\\
    \label{eq:transition_matrix_form}\bs{F} &= \begin{pmatrix}
        1 - \Delta t/\tauc & \Delta t / \tauc \\
        \Delta t / \taus & 1 - \Delta t / \taus
    \end{pmatrix},\\
   \label{eq:discrete:torques}\torques &= \begin{pmatrix}
       \Nc\Delta t/\Ic\\
       \Ns\Delta t/\Is
    \end{pmatrix},\\
       \label{eq:discrete:process_noise}\bs{\Delta W} &= \begin{pmatrix}
       \Delta W_{\textrm c, k}\\
       \Delta W_{\textrm c, k}
    \end{pmatrix}.
    \end{align}
We also define the process noise covariance matrix, $\bs Q$, which is given by
    \begin{align}
    \label{eq:discrete:Q_matrix}\bs{Q} &=\langle\Delta\bs{W}_k\Delta\bs{W}_{k'}^T\rangle \\
    \label{eq:discrete:covar_full}&=\Delta t\,\delta_{k,k'}\begin{pmatrix}
        \sigma_{\rm c}^2/\Ic^2 & 0\\
        0 & \sigma_{\rm s}^2/\Is^2   \end{pmatrix}.
\end{align}
In (\ref{eq:inference:state_transition_equation})--(\ref{eq:discrete:covar_full}), $\bs{x}_k$ is the state vector, $\bs{F}$ is the transition matrix, $\torques$ is sometimes called the control vector, and $\bs Q$ describes the covariance of $\Delta\bs{W}_k$, which has independent increments. 

In general, we have some set of observations $\bs{y}$ with measurement noise $\bs{u}$. We assume that $\bs{y}$ is linearly related to $\bs{x}$ through
\begin{align}
\label{eq:inference:measurement_from_states}
    \bs{y}_k = \bs{B}\bs{x}_k + \bs{u}_k.
\end{align}
For this study, we assume that $\bs{B}$ is the identity matrix, and $\bs{u}$ is Gaussian distributed with covariance matrix, 
\begin{align}
\bs{R} = \langle \bs u_k \bs u_{k'}^T\rangle,
\end{align} which we assume to be diagonal and known \textit{a priori}. However, it is straightforward to relax these assumptions and estimate elements of $\bs R$ and $\bs B$ as well~\citep{Shumway1982,GIBSON20051667}.

\subsection{Inverting the data: an overview of the Kalman filter}
\label{ssec:kalman_filter}
A Kalman filter or smoother efficiently tracks a set of state variables $\bxk$, through time~\citep{kalman1960}. It achieves this by estimating the expectation value and covariance of the state variables at each time step, which we denote as
\begin{align}
\bs{\hat{x}}_{k}&=E(\bs x_k| \bs y, \bs\theta)\label{eq:expectation_of_x_original}\\
\bs\Sigma_{k, k'}&=E[(\bs x_k - \bs{\hat{x}}_{k})(\bs x_{k'} - \bs{\hat{x}}_{k'})^T| \bs y, \bs\theta],\label{eq:covariance_of_x_original}
\end{align}
given a set of observations $\bs y_k$, and a choice of model parameters. We denote the model parameters, in this case the elements of $\transition$, $\torques$ and $\processcovar$, by the vector $\bs\theta$. The state estimation step is necessary because $\bxk$ cannot be inferred from $\bs y_k$ directly due to the random measurement noise $\bs u_k$, and $\bs x_{k+1}$ cannot be inferred from $\bs x_{k}$ directly due to the random (physical) process noise $\bs{\Delta W}_k$.

The Kalman filter works for any linear dynamical system, whenever we can write down equations like (\ref{eq:inference:state_transition_equation}) and (\ref{eq:inference:measurement_from_states}) to estimate $\hbxk$ given $\hbxkmo$ and then update it with $\byk$, respectively. In its simplest form, it is a recursive, causal filter. That is, it estimates $\hbxk$ using measurements $\bs{y}_1,\ldots,\bs{y}_k$. It projects the state estimate forward from one time-step to the next via (\ref{eq:inference:state_transition_equation}), and then updates the projected state estimates using the measurement at that time step. The update is done in a way that minimizes the trace of the state estimate covariance matrix, $\Sigma_{k,k}$.

In this paper, we employ a Rauch-Tung-Streibel (RTS) smoother~\citep{rts_smoother_paper}, which is acausal, meaning we find $\hbxk$ and $\Sigma_{k,k}$ at the $k$-th time-step using all of the available measurements, $\bs{y}_1,\ldots\bs{y}_{N_t}$. This involves running the Kalman filter forward, and then employing a set of backwards recursions to improve $\hbxk$ and further reduce $\textrm{Tr}~\Sigma_{k,k}$.  Once the smoothed estimates, $\hbxk$ and $\Sigma_{k,k}$ are in hand, one can ``invert'' them to obtain a new set of model parameters, $\bs \theta_{\rm new}$. When we run the RTS smoother again with this new set of model parameters the updated state estimates are, in a statistical sense, better than the state estimates we obtained running the smoother with $\bs\theta$. We discuss this technique, known as the expectation-maximization algorithm, in Section~\ref{ssec:infer:em_algorithm}.

\subsection{Expectation-maximization algorithm}
\label{ssec:infer:em_algorithm}
\subsubsection{Overview}
The measurements, $\bs{y}$, can be used to solve for $\tauc$ and $\taus$ in $\bs{F}$, $\Nc$ and $\Ns$ in $\torques$, and $\sigma_{\rm c}$ and $\sigma_{\rm s}$ in $\bs Q$ in (\ref{eq:inference:state_transition_equation})--(\ref{eq:discrete:covar_full}) using any number of maximum-likelihood techniques. In this paper, we primarily discuss the expectation-maximization algorithm~\citep{em_original_paper,Shumway1982}, which naturally extends to situations where observations are missing or $\Delta t$ is not constant~\citep{Shumway1982,Digalakis_1993_EM_algorithm,GIBSON20051667}.
The method cleanly separates \textit{process noise} in the physical variables, characterized by $\bs Q$, and \textit{measurement noise}, characterized by $\bs R$.
Two other common techniques include using Newton's method to maximize the log-likelihood, $\log L$, and the score vector method~\citep{durbin2012time}. The former method relies on numerical estimates of the inverse of the Hessian matrix, while the latter method calculates the gradient of $\log L$ with respect to the parameters and steps towards a maximum of $\log L$~\citep{durbin2012time}.

In this paper, we estimate the elements of $\transition$, $\processcovar$ and $\torques$, as opposed to $\tauc$, $\taus$, $\Nc$, $\Ns$, $\sigma_c$ and $\sigma_s$. It is possible to extend this method to directly estimating the parameters in (\ref{EQ:CRUST_EQ_OF_MOTION}) and (\ref{EQ:SUPERFLUID_EQ_OF_MOTION}) using ``grey-box estimation''~\citep{Ljung:1986:SIT:21413,Ljung2013,KRISTENSEN2004225}.

\subsubsection{Algorithm and its logical basis}

The expectation-maximization algorithm is used to find stationary points of the posterior distribution $p(\bs\theta | \bs y)$ in the absence of a direct observation of the states, $\bs x$. As in the previous section, we use $\bs\theta$ as compact notation for the elements of $\transition$, $\torques$ and $\processcovar$. We use the product rule to separate the posterior in terms of the states
\begin{align}
\log p(\bs\theta | \bs y) = \log p(\bs \theta, \bs x | \bs y) - \log p(\bs x | \bs \theta, \bs y).
\label{eq:likelihood_product_expansion}
\end{align}
The first term on the right-hand side of (\ref{eq:likelihood_product_expansion}) is a multivariate Gaussian with a straightforward interpretation which we discuss in Appendix~\ref{sec:em_overview_for_our_case}. We will use this term to find stationary points of $p(\bs\theta | \bs y)$. 

To start, we treat $\bs x$ as a random variable and take an expectation value of both sides, assuming some fixed value for the parameters, $\bs\theta = \bs\theta_p$, leaving 
\begin{align}
\log p(\bs\theta | \bs y) = E_{p}[\log p(\bs \theta, \bs x | \bs y)] - E_p[\log p(\bs x | \bs \theta, \bs y)]
\label{eq:expansion_of_posterior_distribution}
\end{align}
with
\begin{align}
E_p[\log p(\bs \theta, \bs x | \bs y)] = \int d\bs x \log p(\bs \theta, \bs x | \bs y) p(\bs x | \bs \theta_p, \bs y).
\label{eq:expectation_of_posterior}
\end{align}
The second term on the right-hand-side of (\ref{eq:expansion_of_posterior_distribution}) is maximized at $\bs\theta = \bs\theta_p$~\citep{em_original_paper,gelman2013bayesian}, while the first term on the right-hand-side can be maximized analytically by a new choice of $\bs\theta$, which increases the marginal posterior distribution on the left-hand-side of (\ref{eq:expansion_of_posterior_distribution}). This lends itself to the following iterative algorithm (where $p$ now defines how many times we have iterated):
\begin{enumerate}[leftmargin=3ex]
	\item Start with a guess of $\bs\theta=\bs\theta_0$.
	\item Calculate $E_p[\log p(\bs \theta, \bs x | \bs y)]$ (``E-step'') using current estimate of the parameters $\bs\theta_p$.
	\item Maximize $E_p[\log p(\bs \theta, \bs x | \bs y)]$ from step (ii) over $\bs\theta$ (``M-step''). Set $\bs\theta_{p+1}$ to the maximum.
	\item Repeat steps (ii) and (iii) for a fixed number of iterations or until $\log p(\bs\theta | \bs y)$ converges
\end{enumerate}

In Appendix~\ref{sec:em_overview_for_our_case} we discuss the application of this method to our specific case. We first present the full form of $p(\bm\theta,\bm x| \bm y)$. We then discuss the E-step and M-step, along with the implementation used in carrying out the simulations discussed in Section~\ref{sec:simulated_data_results}. In Appendix~\ref{app:em_implementation} we show the detailed formulae used in carrying out the E-step and the M-step. We calculate~(\ref{eq:expectation_of_posterior}) explicitly, and show how to find the values of the parameters, $\bs\theta$, that maximize~(\ref{eq:expectation_of_posterior}). We finish Appendix~\ref{app:em_implementation} by showing the steps in the RTS smoother used to evaluate state-variable estimates.

\section{Tests With Synthetic Data}
\label{sec:simulated_data_results}
In this section we use synthetic data to test the recovery of parameters in the two-component model in (\ref{EQ:CRUST_EQ_OF_MOTION}) and (\ref{EQ:SUPERFLUID_EQ_OF_MOTION}). 
We discuss how we simulate the data in Section~\ref{ssec:results:generate_data} and justify the illustrative choices shown in Table~\ref{tab:input_params}. We then estimate the injected parameters under two scenarios. In Section~\ref{ssec:results:gw_and_em_data} we analyse hypothetical gravitational-wave and electromagnetic measurements of $\omgs(t)$ and $\omgc(t)$. In Section~\ref{ssec:results:em_only_data} we analyse hypothetical electromagnetic measurements of $\omgc(t)$.
In both scenarios, we take $\bs R$ to be diagonal with entries given by $10^{-18}~\rm{rad^2~s^{-2}}$.

To characterize the algorithm, we run~\Nsimulations~realizations of process noise and measurement noise with the parameters given in Table~\ref{tab:input_params}. We then estimate the parameters of each realization independently using the algorithm presented in Section~\ref{ssec:infer:em_algorithm}. The method we use is as follows:
\begin{enumerate}[(a),leftmargin=3ex]
    \item Generate a realization of synthetic data. 
    \item Estimate parameters using 100 random initial state and parameter choices. Of those 100 estimates, choose the one with the largest likelihood.
    \item Repeat steps (a) and (b) \Nsimulations~times.
\end{enumerate}
Step (b) is necessary because the algorithm converges to a local maximum. Therefore, we start the algorithm in 100 randomly chosen locations, record where the algorithm finishes for each location, and then take the estimate that maximizes the log-likelihood. We assume that this is a reasonable approximation to the global maximum of the log-likelihood. The distribution of starting points for each parameter we consider is shown in Table~\ref{tab:starting_points}.

Using this method to characterize our algorithm, we then compare the estimated and injected parameters for each realization. In the scenario where we have gravitational-wave and electromagnetic measurements we accurately estimate all parameters in the model. When we only have electromagnetic measurements, the dynamics of the superfluid are observed indirectly through interaction with the crust, and some parameters and combinations of parameters, are estimated accurately, while others are not.

\subsection{Data}
\label{ssec:results:generate_data}
We generate the synthetic time-series $\omgs(t_0),\ldots,\omgs(t_{N_t})$ and $\omgc(t_0),\ldots,\omgc(t_{N_t})$ by integrating (\ref{EQ:CRUST_EQ_OF_MOTION}) and (\ref{EQ:SUPERFLUID_EQ_OF_MOTION}) numerically
with a stochastic Runge-Kutta It\^o integration technique~\citep{10.2307/41062628}, using the implementation in the  
\texttt{sdeint} python package\footnote{\url{https://github.com/mattja/sdeint}}. 
We use $\Delta t = t_{k+1} - t_k = 86400~\textrm{s}$ and we generate 1157 time steps. 
The parameter values shown in Table~\ref{tab:input_params} are meant to be illustrative. The timing noise induced by $\sigma_{\rm c} / \Ic$ and $\sigma_{\rm s} / \Is$ is comparable to the largest timing noise observed in radio pulsars by~\cite{2010MNRAS.402.1027H}, i.e. $\log_{10}\sigma_z(10\textrm{ yr})\sim -6.5$, where $\sigma_z$ is defined in~\cite{matsakis:1997lsd}. It also falls within the levels of timing noise reported for magnetars and X-ray pulsars by~\citet{10.1093/mnras/sty3213}. In that case, we find $\log_{10}{S_r} \sim -18$ where $\log_{10}{S_r}$ is defined in~\cite{10.1093/mnras/sty3213} and has a range of $(-24, -16)$ for the set of magnetars and X-ray pulsars considered. We present a quantitative comparison between our timing noise statistics, $\sigma_{\rm c} / \Ic$ and $\sigma_{\rm s} / \Is$, and $S_r$ in Appendix~\ref{sec:timing_noise_comparison}, including a discussion of how the relaxation process affects these timing noise estimates.

\begin{figure*}
    \centering
    \includegraphics[width=0.7\textwidth]{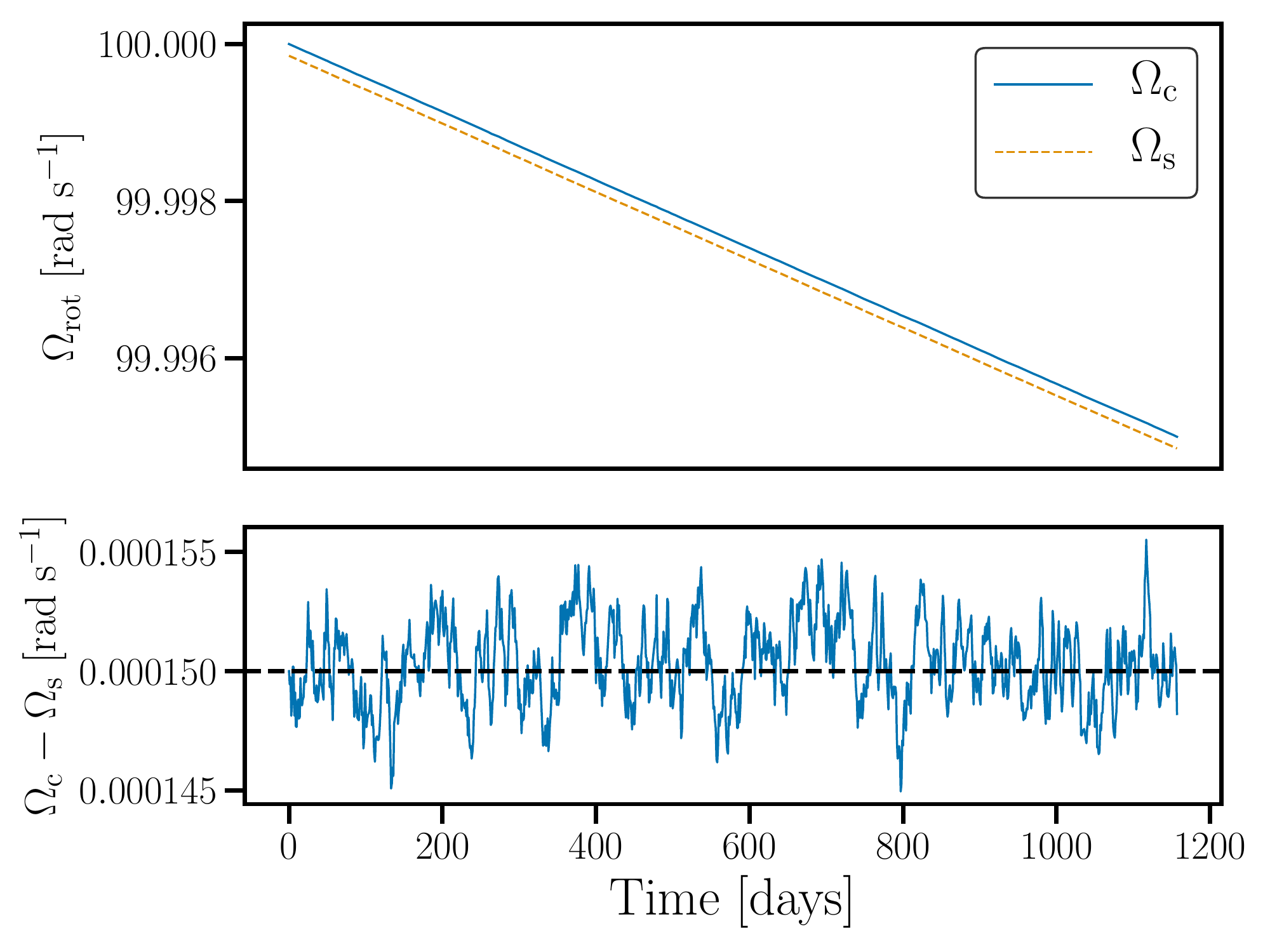}
    \caption{Simulated data generated by numerically integrating~(\ref{EQ:CRUST_EQ_OF_MOTION}) and (\ref{EQ:SUPERFLUID_EQ_OF_MOTION}) with the parameters shown in Table~\ref{tab:input_params}. The top panel shows the angular velocity of the crust (blue solid curve) and the superfluid (orange dashed curve). The bottom panel shows the measured lag between the angular velocities of the two components (blue solid curve) and the expected lag (black dashed curve) predicted analytically by~(\ref{eq:simple_model:lag_limit_long_time}).}
    \label{fig:simulated_data}
\end{figure*}

A sample of synthetic data is shown in Fig.~\ref{fig:simulated_data}. 
The top panel shows $\omgs(t)$ and $\omgc(t)$. Both components spin down secularly as well as wander. The bottom panel shows the difference, $\omgc(t)-\omgs(t)$,
which varies stochastically around the expected ensemble-averaged value (indicated by the dashed black line), calculated using  
(\ref{eq:simple_model:lag_limit_long_time}).
\begin{table}
{\renewcommand{\arraystretch}{1.2}
\begin{center}
\begin{tabular}{l l l}
\hline
    {\bf Parameter} & {\bf Value} & {\bf Units}\\
    \hline\hline
    $\Nc/\Ic$ & $10^{-10}$ & $\rm{rad~s^{-2}}$ \\
    $\Ns/\Is$ & $-10^{-10}$ & $\rm{rad~s^{-2}}$ \\
    $\tauc$ & $10^{6}$ & s \\
    $\taus$ & $3\times10^{6}$ & s \\
    $\sigma_{\rm c} / I_c$ & $2.5 \times 10^{-9}$ & $\rm{rad~s^{-3/2}}$\\
    $\sigma_{\rm s} / I_s$ & $1.25\times 10^{-9}$ & $\rm{rad~s^{-3/2}}$\\
    $\Omega_{\rm c}(0)$ & 100 & $\rm{rad~s^{-1}}$\\
    $\Omega_{\rm s}(0)$ & 99.99985 & $\rm{rad~s^{-1}}$\\
    $t_{k+1} - t_k$ & $86400$ & s \\
    \hline
\end{tabular}
\caption{Parameters used to create simulated data discussed in Section~\ref{ssec:results:generate_data} and shown in Fig.~\ref{fig:simulated_data}. $\omgs(0)$ is chosen such that $\omgc(0) - \omgs(0)$ takes the expected value in~(\ref{eq:simple_model:lag_limit_long_time}). Otherwise there is a transient move towards this value at early time steps. The choice $\Nc/\Ic>0$ simulates a system that is accreting.}
\label{tab:input_params}
\end{center}
}
\end{table}

\subsection{Electromagnetic and gravitational-wave measurements}
\label{ssec:results:gw_and_em_data}
Let us begin with the hitherto hypothetical situation where both electromagnetic and gravitational-wave timing data are available. We assume that the electromagnetic and gravitational-wave measurements correspond directly to measurements of $\omgc(t)$ and $\omgs(t)$ respectively, i.e. that $\bs B$ in (\ref{eq:inference:measurement_from_states}) is the identity matrix. It is straightforward to generalize to other scenarios too, e.g. where the gravitational-wave data correspond to a linear combination of $\omgc$ and $\omgs$. We include a constraint to insist that summing across the columns of $\bs F$ yields unity, a constraint that is clear from (\ref{eq:transition_matrix_form}). As described in the introduction to Section~\ref{sec:simulated_data_results}, we start the algorithm at 100 randomly-generated starting points for each noise realization. The prior distributions from which the starting points are drawn are given in Table~\ref{tab:starting_points}. For the relaxation time-scales, this corresponds to an allowable range between roughly three days and three years, which is consistent with the range of time-scales seen in pulsar glitch relaxations~\citep{2010MNRAS.409.1253V}. The noise amplitudes, $\sigma_{\rm c}/\Ic$ and $\sigma_{\rm s}/\Is$, are consistent with $S_r$ in~\citet{10.1093/mnras/sty3213}, as discussed previously. The magnitudes of the torques, $\Nc/\Ic$ and $\Ns/\Is$, encompass a range of typical spin-down rates observed for pulsars.

\begin{table}
{\renewcommand{\arraystretch}{1.2}
\begin{center}
\begin{tabular}{l l l}
\hline
    {\bf Parameter} & {\bf Distribution of starting points}\\
    \hline\hline
$\Delta t / \tauc$   & $\textrm{LogUniform}(0.001, 0.3)$\\
$\Delta t / \taus$   & $\textrm{LogUniform}(0.001, 0.3)$\\
$\sigma_{\rm c}/\Ic~[\rm{rad~s^{-3/2}}]$ & $\textrm{LogUniform}(10^{-12}, 10^{-7})~$\\
$\sigma_{\rm s}/\Is~[\rm{rad~s^{-3/2}}]$ & $\textrm{LogUniform}(10^{-12}, 10^{-7})$\\
$\Nc/\Ic~[\rm{rad~s^{-2}}]$		   & $\textrm{LogUniform}(-10^{-15}, -10^{-5})$\\
$\Ns/\Is~[\rm{rad~s^{-2}}]$           & $\textrm{LogUniform}(10^{-15}, 10^{-5})$\\
    \hline
\end{tabular}
\caption{Distributions from which the starting parameters for the algorithm described in Section~\ref{ssec:infer:em_algorithm} are drawn when running on synthetic data.}
\label{tab:starting_points}
\end{center}
}
\end{table}

\begin{figure*}
    \includegraphics[width=0.7\textwidth]{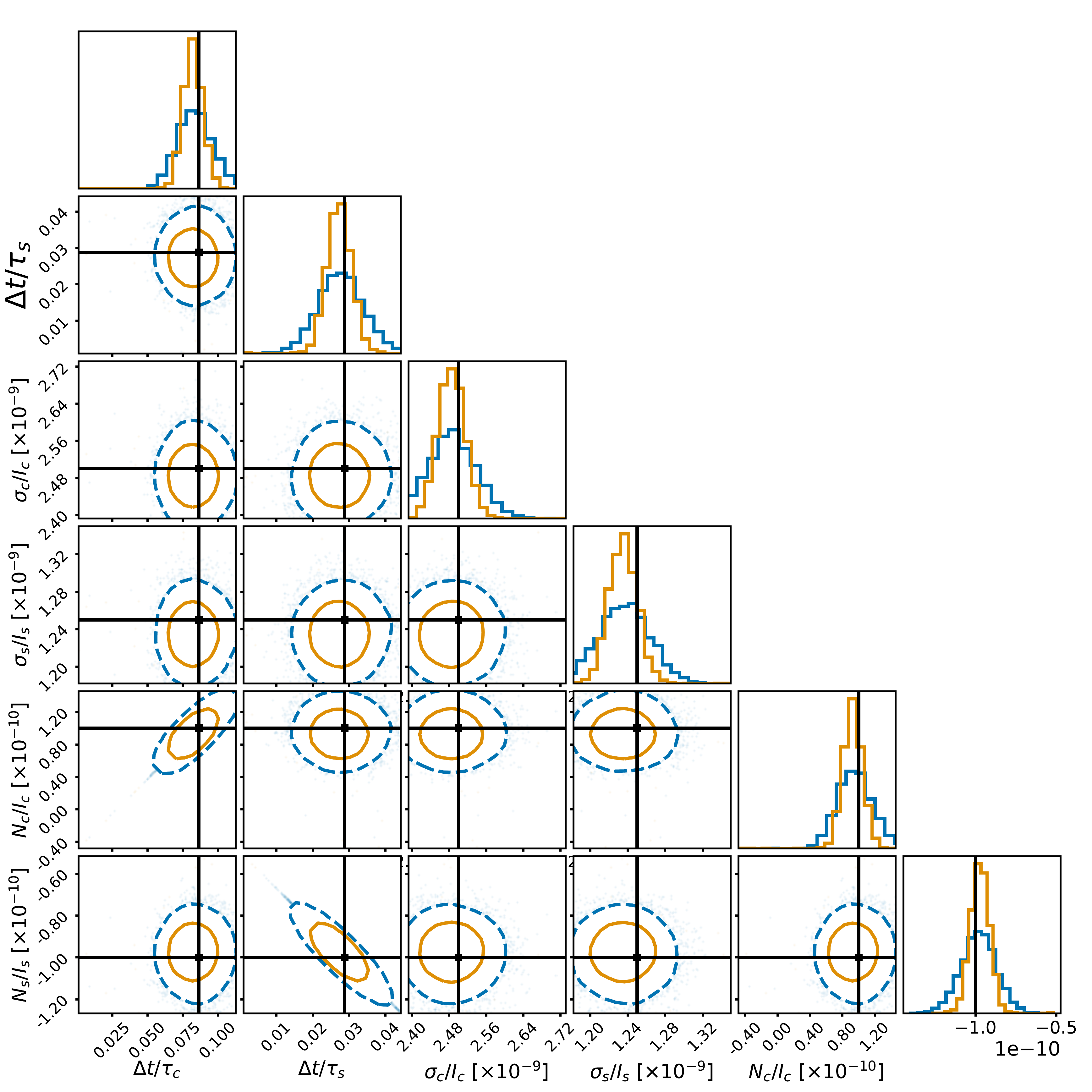}
    \caption{Estimates of $\Delta t /\tauc$, $\Delta t/ \taus$, $\sigma_{\rm c} / \Ic$, $\sigma_{\rm s} / \Is$, $\Nc/\Ic$ and $\Ns/\Is$ using electromagnetic and gravitational wave measurements for \Nsimulations~realizations (units are given in Table~\ref{tab:input_params}). Each realization uses the same input parameters, but different realizations of process and measurement noise. We show 90\% contours for $N_t=1157$ (blue dashed curve) and $N_t=4630$ (orange solid curve). We see that the estimates are clustered around the true values (indicated with the intersecting black dashed lines). The figure is presented in the traditional format of a ``corner plot,'' but the reader is reminded that this is merely a presentational device; the plotted results are not samples drawn from a Markov chain Monte Carlo analysis.}
    \label{fig:em_gw_measurements:corner_plot}
\end{figure*}

We show a corner plot for estimates of the non-zero elements of $\transition$, $\torques$, and $\processcovar$,  in Fig.~\ref{fig:em_gw_measurements:corner_plot}. Corner plots are often used to display Markov chain Monte Carlo posteriors, but here they have a different meaning: each point shows the parameter estimates for one of the 5000 different noise realizations. The contours in Fig.~\ref{fig:em_gw_measurements:corner_plot} indicate boundaries within which 90\% of the estimates lie. We would also like to see whether we are able to increase the accuracy of our measurements by using longer data sets. To this end, we present results with $N_t = 1157$ (dashed blue curves) and $N_t=4630$ time-steps (orange solid contours). The fact that the solid orange contours surround a smaller region of parameter space indicate that the parameter estimates are better when more data are included. The estimated parameters cluster around the true values for each parameter in Table~\ref{tab:input_params}, as indicated by the intersecting black dotted lines.

In Table~\ref{tab:estimated_params} we present estimates of $\tauc$, $\taus$, $\sigma_c/\Ic$, $\sigma_s/\Is$, $\Nc/\Ic$ and $\Ns/\Is$. The reported values are the median of the~\Nsimulations~estimates and the quoted ranges are the 5th and 95th percentiles of the~\Nsimulations~parameter estimates. In the case of $\tauc$, we convert each estimate of $\Delta t/\tauc$ to $\tauc$ before estimating the median and the range, and likewise for $\taus$. 

How do the estimated parameters compare to the true values?
The estimated recovery time-scales are $\taus = 3.13^{+1.81}_{-0.89}\times 10^6~\rm{s}$ and $\tauc = 1.04^{+0.30}_{-0.22}\times10^6~{\rm s}$ for $N_t=1157$. The ranges include the injected value shown in Table~\ref{tab:input_params}.
The estimates of $\Nc/\Ic$ and $\Ns/\Is$ are also clustered around the injected values, but
the diagonal ellipse in the $\Ns/\Is$ vs. $\Delta t /\taus $ contour plot indicates that these parameters are degenerate with one another (likewise for $\Delta t / \tauc$ and $\Nc/\Ic$). These degeneracies can be understood in (\ref{eq:deterministic_spindown}), where $\Ns/\Is$ and $\Nc/\Ic$ only enter in terms where they are multiplied by $\taus$ and $\tauc$ respectively.

\begin{table*}
\begin{center}
{\renewcommand{\arraystretch}{1.5}
\begin{tabular}{l | c | c | c | c | c|}
\hline
    {\bf Parameter}                                        & {\bf True}& {\bf GW/EM short}& {\bf GW/EM long}& {\bf EM short} & {\bf EM long}\\
    \hline
    $\tauc \times 10^{-6}$ [s]                             & $1$ & $1.04^{+0.30}_{-0.22}$   & $1.05^{+0.15}_{-0.13}$  & $ 1.19^{+85.20}_{-0.77}$ & $1.27^{+0.76}_{-0.71}$\\
    $\taus\times10^{-6}$ [s]                               & $3$ & $3.13^{+1.81}_{-0.89}$   & $3.17^{+0.80}_{-0.53}$  & $ 1.34^{+46.99}_{-1.05}$  & $1.92^{+18.37}_{-1.38}$ \\
    $\Nc/\Ic\times 10^{10}$  [$\rm{rad~s^{-2}}$]           & $1$ & $0.95^{+0.39}_{-0.32}$  & $0.92^{+0.20}_{-0.18}$   & $-0.50^{+9.96}_{-9.53}$ & $-0.25^{+8.40}_{-8.65}$\\
    $\Ns/\Is\times 10^{10}$  [$\rm{rad~s^{-2}}$]         & $-1$& $-0.98^{+0.17}_{-0.19}$ & $-0.97^{+0.10}_{-0.09}$ & $-0.50^{+19.72}_{-21.09}$ & $-0.59^{+7.66}_{-7.43}$\\
    $\sigma_{\rm c} / \Ic\times10^{9}$ [$\rm{rad~s^{-3/2}}$] & $2.5$ & $2.49^{+0.09}_{-0.09}$ & $2.48^{+0.05}_{-0.05}$& $2.50^{+0.10}_{-0.16}$  & $ 2.50^{+0.06}_{-0.10}$\\
    $\sigma_{\rm s} / \Is\times10^{9}$ [$\rm{rad~s^{-3/2}}$] & $1.25$ & $1.24^{+0.04}_{-0.04}$ & $1.23^{+0.02}_{-0.02}$ & $0.96^{+27.39}_{-0.96}$ & $0.67^{+1.12}_{-0.67}$\\
    \hline
\end{tabular}
\caption{Estimated parameters with error ranges for all scenarios we considered. In column ``True'' we show the input parameters from Table~\ref{tab:input_params}. In column ``GW/EM  short'' we show median and 90\% range of recovered estimates for $N_t = 1157$ using gravitational wave and electromagnetic data. In column ``GW/EM long'' we show the same but for $N_t=4630$. The final two columns are analogous but for electromagnetic only data. We see that for the gravitational wave and electromagnetic results the true values fall within the inferred range. However, for the electromagnetic only case, $\Nc/\Ic$, $\Ns/\Is$, $\tau_s$ and $\sigma_s$ all show a bias in the estimates, as discussed further in Section~\ref{ssec:results:em_only_data}.}
\label{tab:estimated_params}
}
\end{center}
\end{table*}

\subsection{Electromagnetic measurements only: $\Nc/\Ic$, $\Ns/\Is$, $\sigma_{\rm c}/\Ic$, $\sigma_{\rm s}/\Is$}
\label{ssec:results:em_only_data}

As continuous gravitational waves from a rapidly rotating neutron star have not been detected yet, it is worth asking how accurately we are able to estimate the parameters of the two-component model using electromagnetic pulse timing data only. We assume we have a time-series measurement of the crust angular velocity, $\omgc(t)$, so that $y_k$ is now a scalar, as opposed to a column vector, and $\bs B = (1, 0)$ is now a row vector instead of a 2x2 matrix.

\begin{figure*}
    \includegraphics[width=0.7\textwidth]{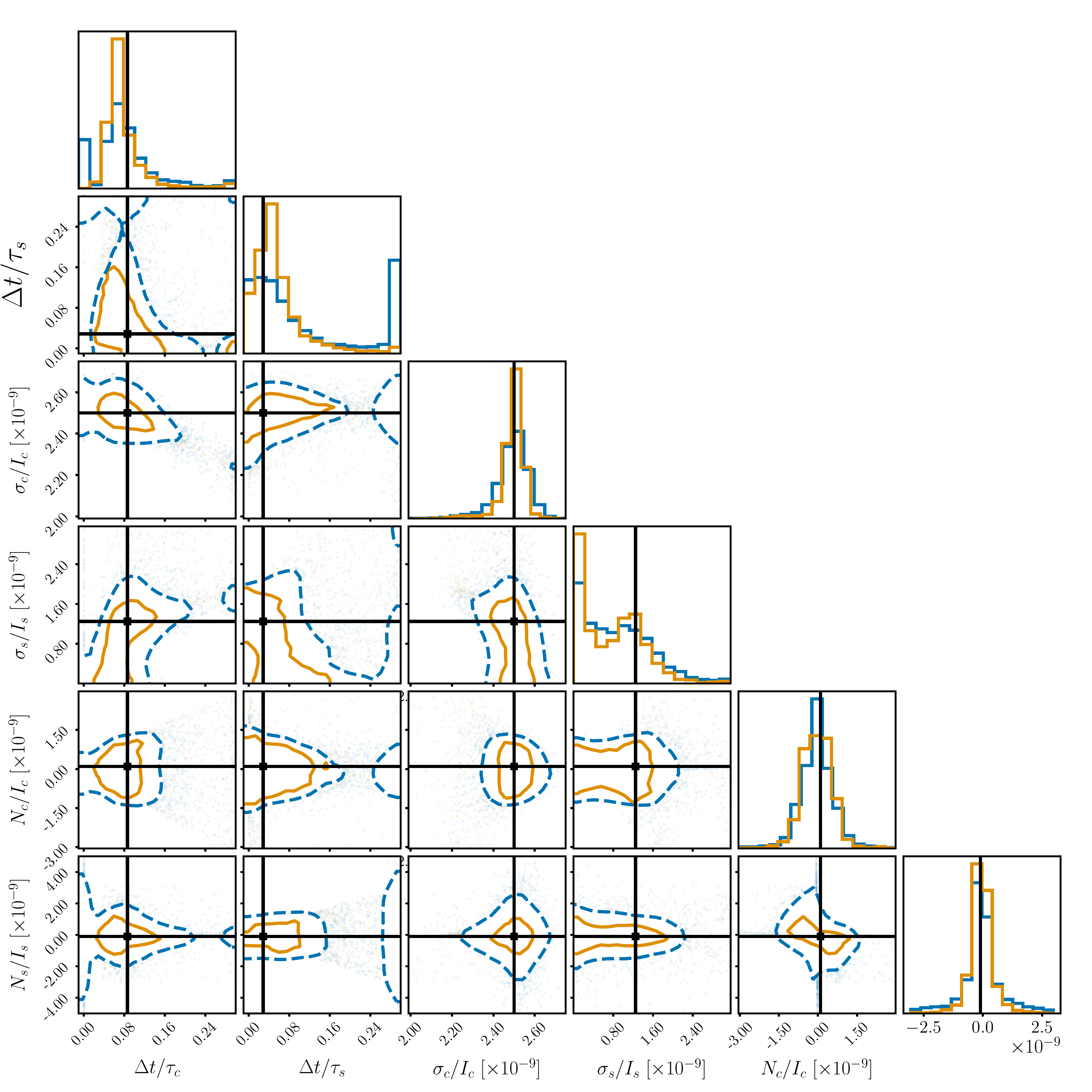}
    \caption{Estimates of $\Delta t /\tauc$, $\Delta t/ \taus$, $\sigma_{\rm c} / \Ic$, $\sigma_{\rm s} / \Is$, $\Nc/\Ic$ and $\Ns/\Is$ using electromagnetic only measurements for \Nsimulations~realizations (units given in Table~\ref{tab:input_params}). Each realization uses the same input parameters, but different realizations of process and measurement noise. We show 90\% contours for $N_t=1157$ (blue dashed curve) and $N_t=4630$ (orange solid curve). We see that the estimates are clustered around the true values (indicated with the intersecting black dashed lines) for $\tauc$ and $\sigma_{\rm c}$, but there is a bias in the estimates of $\taus$, $\Nc/\Ic$, $\Ns/\Is$ and $\sigma_{\rm s}$. We discuss this further in Section~\ref{ssec:results:em_only_data}. The figure is presented in the traditional format of a ``corner plot,'' but the reader is reminded that this is merely a presentational device; the plotted results are not samples drawn from a Markov chain Monte Carlo analysis..}
    \label{fig:em_only_measurements:corner_plot}
\end{figure*}

The results of \Nsimulations~estimates, each with different realizations of process and measurement noise, are shown in Fig.~\ref{fig:em_only_measurements:corner_plot}. We again consider two data sets that are different lengths to test whether including more data improves recovery. The dashed blue contours correspond to the data set with $N_t=1152$. The solid orange contours correspond to the data set with $N_t=4630$. The contours enclose 90\% of the parameter estimates, but it is important to note that these do not represent posterior distributions on the parameters. Rather they represent the dispersion of the estimated maximum likelihood estimates given~\Nsimulations~independent realizations of process and measurement noise, just as in Fig.~\ref{fig:em_gw_measurements:corner_plot}.

In Fig.~\ref{fig:em_only_measurements:corner_plot}, and in the values in the rightmost two columns of Table~\ref{tab:estimated_params}, we show that we are able to accurately estimate $\tauc$ and $\sigma_{\rm c}/\Ic$ for $N_t=4630$. For $N_t=1157$, there are two peaks in the distribution of estimated $\Delta t/\tauc$ and $\Delta t/\taus$, which we discuss in detail below. The noise associated with the crust, $\sigma_c/\Ic$ is well-estimated for both $N_t=1157$ and $N_t=4630$, but the noise on the superfluid $\sigma_{\rm s}/\Is$, is not well-constrained for $N_t=1157$. The range of estimates narrows considerably for $\sigma_{\rm s}/\Is$ moving from  $N_t=1157$ to $N_t=4630$, and there is evidence of a peak emerging near the correct value, indicating that we get more accurate and precise estimates of this parameter as we add more data.  Meanwhile, the median estimates for both $\Nc / \Ic$ and $\Ns / \Is$ have percent errors of $150\%$ and $50\%$ respectively for $N_t=1157$, which improve to $125\%$ and $41\%$ for $N_t=4630$.  As we discuss below, we can accurately estimate the ensemble-averaged spindown $\langle\dot\Omega_{\rm c,s}\rangle$.

\subsection{Electromagnetic measurements only: $\langle\dot\Omega_{\rm c, s}\rangle$, $\tau$, $\langle\omgc(t) - \omgs(t)\rangle$}
\begin{figure}
\begin{center}
	\includegraphics[width=0.4\textwidth]{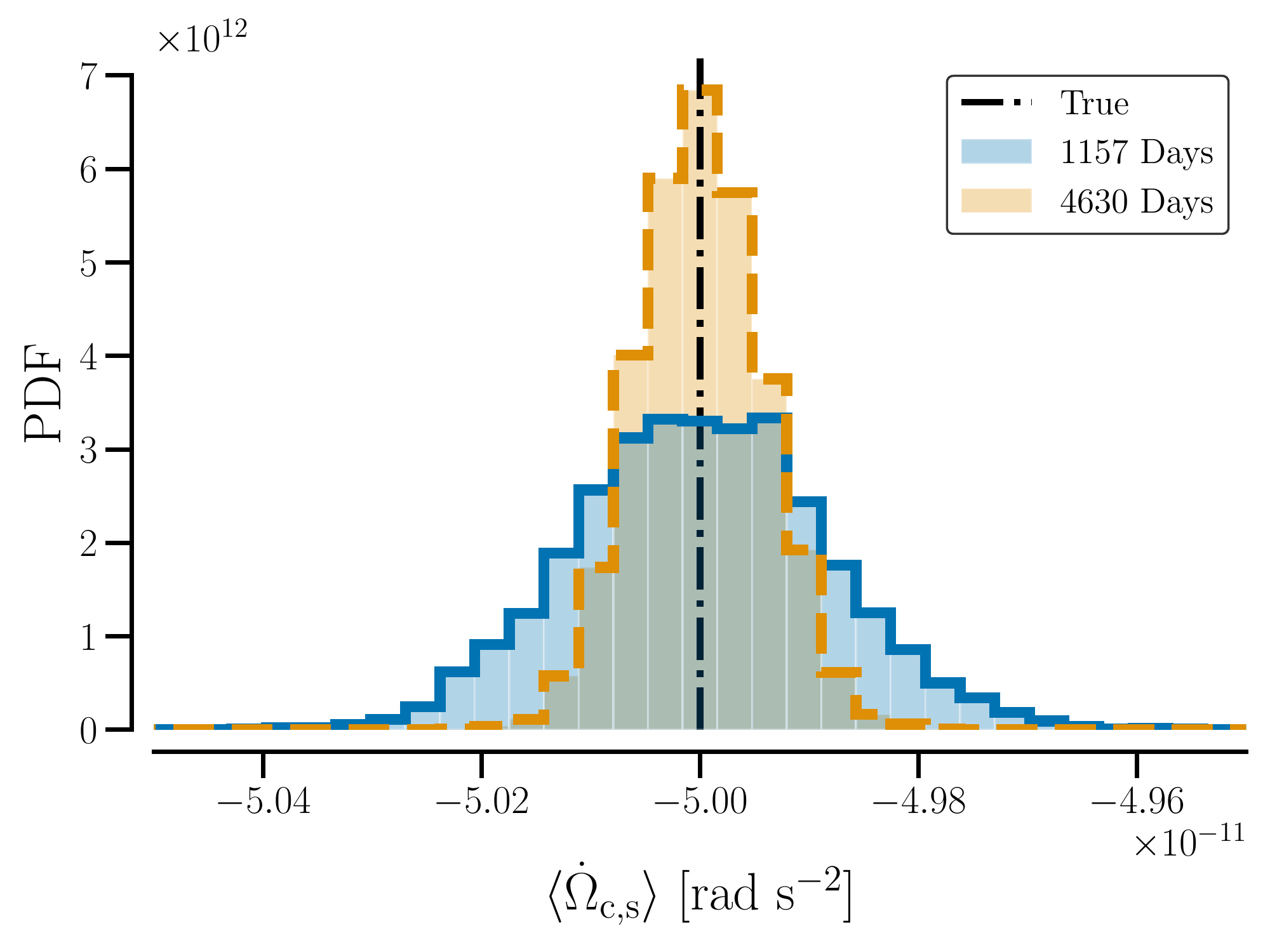}
\includegraphics[width=0.4\textwidth]{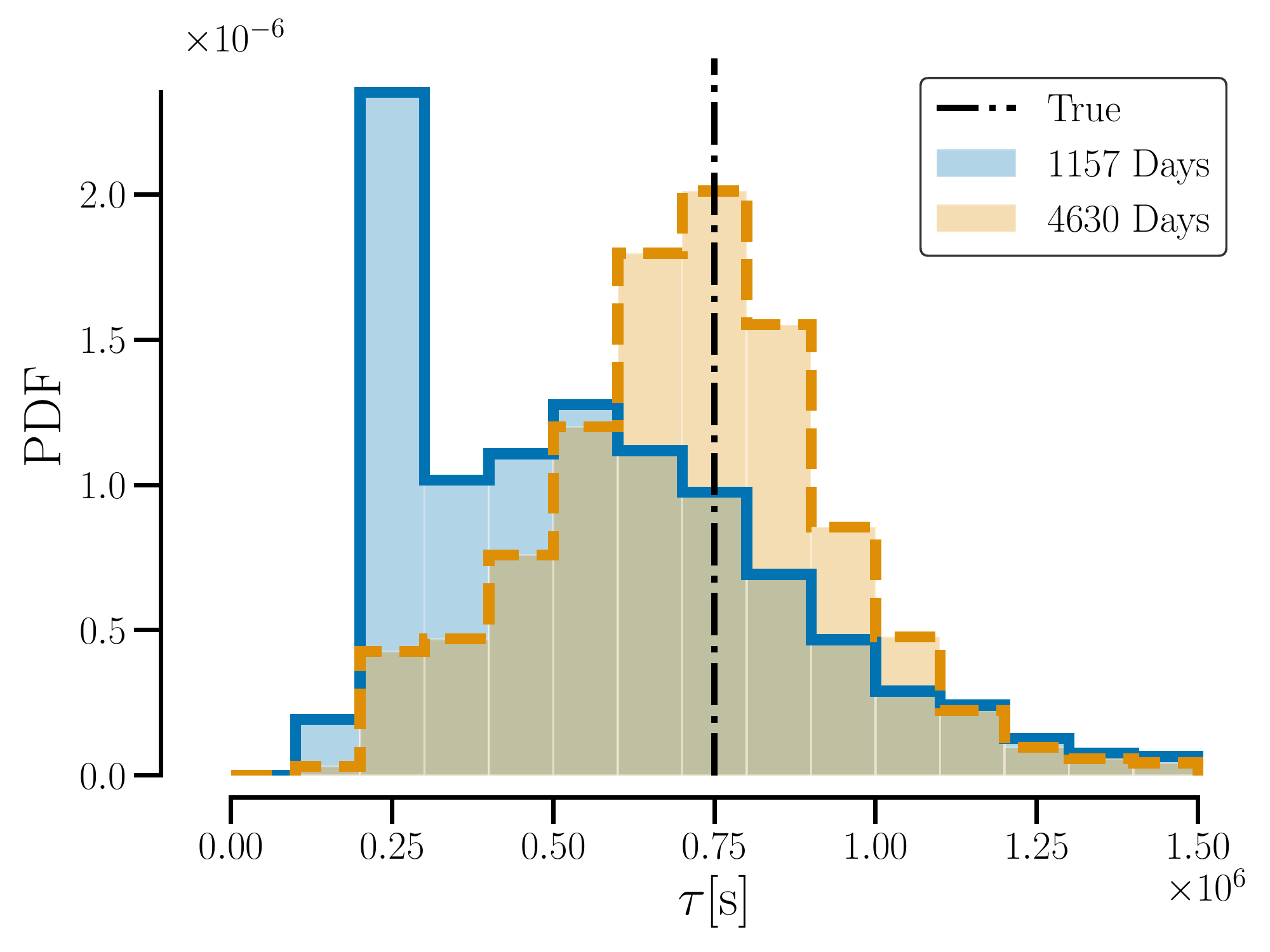}
	\end{center}
	\caption{Probability distribution of $\langle \dot\Omega_{\rm c,s}\rangle$ (top) and $\tau$ (bottom) for the electromagnetic only case. For each simulation we use the estimated model parameters to calculate $\langle\dot\Omega_{\rm c,s}\rangle$ in (\ref{eq:deterministic_spindown}) and $\tau$ in (\ref{eq:tau}), and we present the resulting distributions. The orange, dashed histogram is for $N_t=4630$, while the blue, solid histogram is for $N_t=1157$. The vertical dash-dotted line indicates the expected value based on the injected parameters. We are consistently able to estimate $\langle\dot\Omega_{\rm c,s}\rangle$, even when we are unable to recover $\Nc/\Ic$ and $\Ns/\Is$ as seen Fig.~\ref{fig:em_only_measurements:corner_plot}. We are often able to often estimate $\tau$ correctly for $N_t=1157$. However there is a peak at low values of $\tau$ in the blue bottom histogram, as in Fig.~\ref{fig:nc_tauc_ns_taus}.}
	\label{fig:omega_dot_em_only}
\end{figure}

While we are unable to accurately estimate $\Nc/\Ic$ and $\Ns/\Is$, the ensemble averaged spindown, $\langle\dot\Omega_{\rm c, s}\rangle$, expressed in (\ref{eq:deterministic_spindown}), is well-constrained. If we take the estimated parameters from each of the~\Nsimulations~realizations and combine them using (\ref{eq:deterministic_spindown}), we find the distribution plotted in the top panel of Fig.~\ref{fig:omega_dot_em_only} for $N_t=1157$ (blue, solid) and $N_t=4630$ (orange, dashed). The physical interpretation of this value is the ensemble-average of the derivative of the crust and superfluid angular velocity. The dashed-dotted line in Fig.~\ref{fig:omega_dot_em_only} corresponds to the value of $\langle \dot\Omega_{\rm c,s}\rangle$ calculated using the input parameters in Table~\ref{tab:input_params}. The fact that the distributions peak at the dashed-dotted line indicates that, while estimates of $\Nc/\Ic$ and $\Ns/\Is$ are incorrect, the specific combination in (\ref{eq:deterministic_spindown}) yields the correct ensemble-averaged, spin-down rate of both components, with $\langle\dot\Omega_{\rm c,s}\rangle = -5.00^{+0.01}_{-0.01} \times 10^{-11}~\rm{rad~s^{-2}}$ for $N_t=4630$. In the bottom panel of Fig.~\ref{fig:omega_dot_em_only}, we plot the distribution of $\tau$, defined in (\ref{eq:tau}), calculated for each of the realizations. For $N_t=4630$ the distribution clearly peaks around the injected value, with $\tau = 7.18^{+3.94}_{-4.12}\times 10^5~\rm{s}$, which is a smaller range than for either of the individual relaxation times reported in Table~\ref{tab:estimated_params}. However, for $N_t=1157$, there is a peak at lower values that is related to the bimodal features seen in the first two columns of Fig.~\ref{fig:em_only_measurements:corner_plot}, which are discussed in detail in Section~\ref{ssec:em_only_bimodality}.

\begin{figure}
\includegraphics[width=0.4\textwidth]{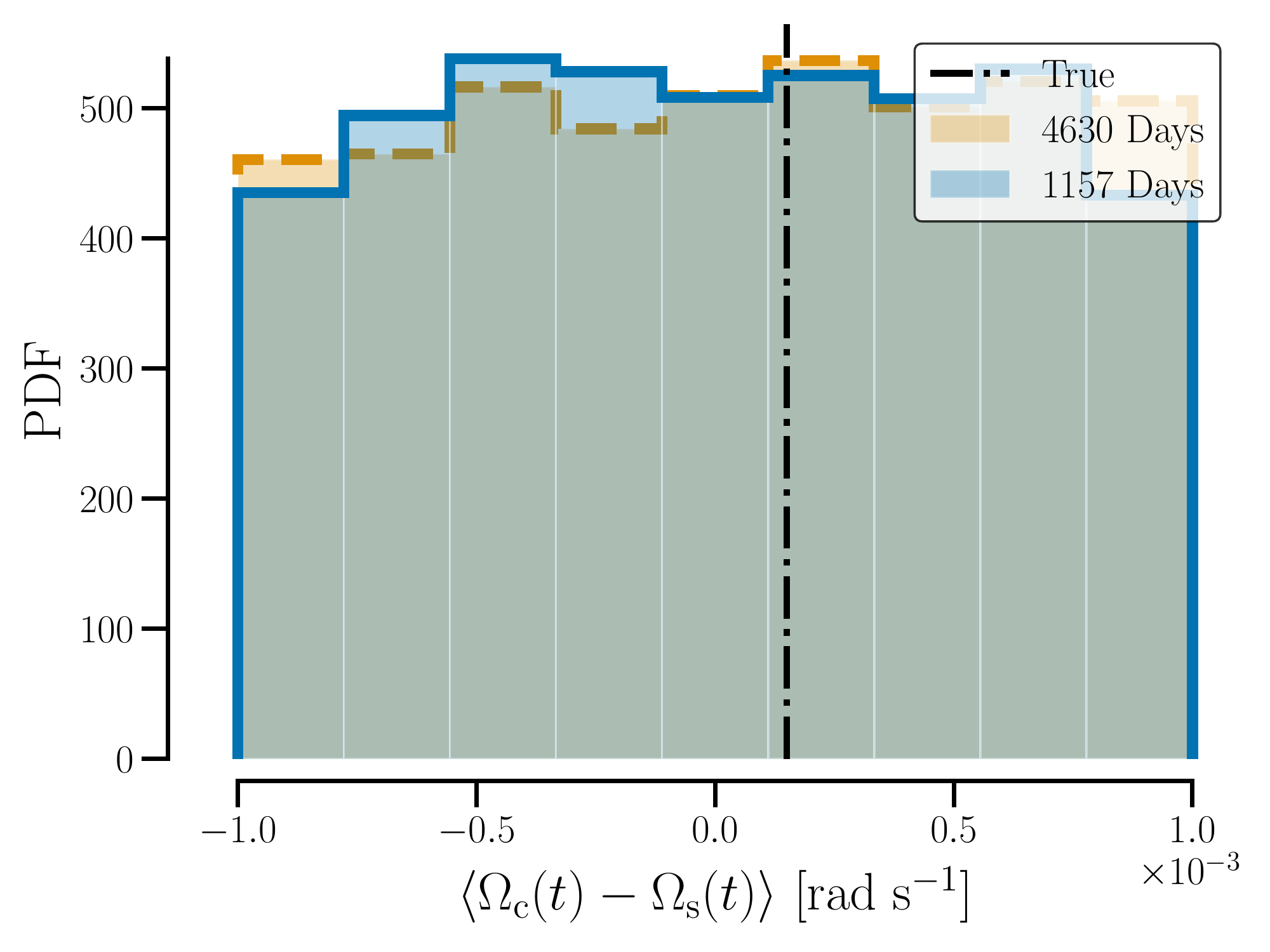}
\caption{Probability distribution of the lag between components, $\langle\omgc(t) - \omgs(t)\rangle$, for the electromagnetic only data. For each simulation we use the estimated model parameters to calculate $\langle\omgc(t) - \omgs(t)\rangle$ as expressed in~(\ref{eq:simple_model:lag_limit_long_time}). The orange, dashed histogram is for $N_t=4630$, while the blue, solid histogram is for $N_t=1157$. The dash-dotted black line indicates the lag calculated using the injected parameters, $\langle\omgc(t) - \omgs(t)\rangle = 1.5\times10^{-4}~\rm{rad~s^{-1}}$. We are unable to accurately estimate the lag in either case. Instead, the lag is uniformly distributed between $-10^{-3}$ and $10^{-3}$, which represents the range of starting points given to the iterative solver.}
\label{fig:lag_plot_em_only}
\end{figure}

We now turn to the ensemble-averaged lag between the crust and the core. Using~(\ref{eq:simple_model:lag_limit_long_time}) and the values in Table~\ref{tab:input_params}, the injected ensemble-averaged lag is $\langle\omgc(t) - \omgs(t)\rangle = 1.5\times10^{-4}~\rm{rad~s^{-1}}$. In Fig.~\ref{fig:lag_plot_em_only} we show the ensemble-averaged lag calculated using parameter estimates for each of the $\Nsimulations$ realizations for $N_t=1157$ (blue, solid) and $N_t=4630$ (orange, dashed). The estimates are uniformly distributed between $-10^{-3}~\rm{rad~s^{-1}}$ and $10^{-3}~\rm{rad~s^{-1}}$, which represent the range of starting points given to the iterative solver, meaning that with only electromagnetic measurements we are unable to estimate the lag between the two components.

\subsection{Bimodality}
\label{ssec:em_only_bimodality}
\begin{figure}
    \centering
    \includegraphics[width=0.4\textwidth]{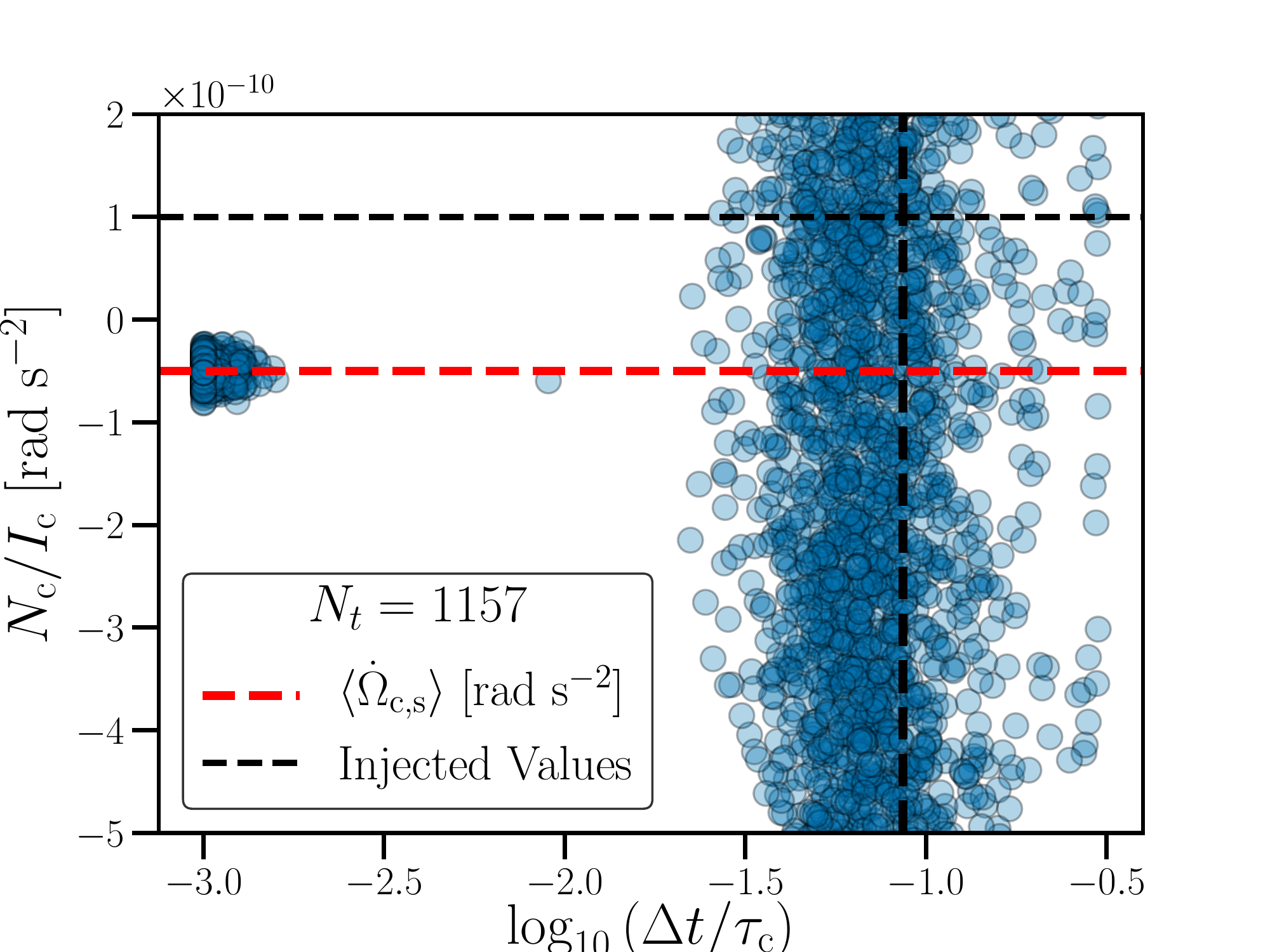}
    \includegraphics[width=0.4\textwidth]{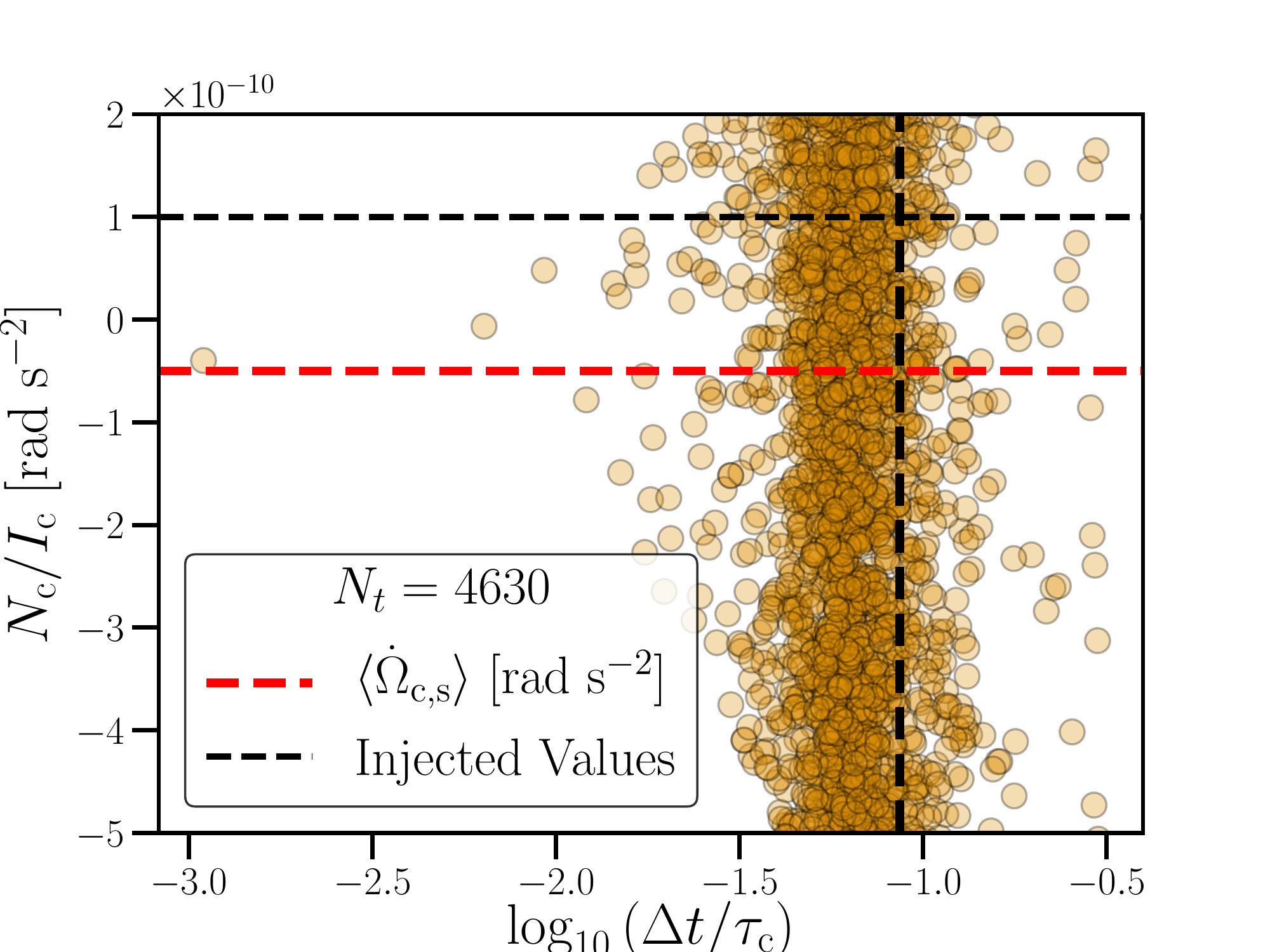}
    \caption{Estimated parameters for $\Delta t/\tauc$ (horizontal axis) and $\Nc/\Ic$ (vertical axis) for each simulation for $N_t=1157$ (top) and $N_t=4630$ (bottom) for the electromagnetic only case. For $N_t=1157$ there is a cluster of points near $\Delta t/\tauc \approx 10^{-3}$ (which is the low end of the allowable range) and $\Nc/\Ic \approx \langle\dot\Omega_{\rm c,s}\rangle$. This cluster of points is consistent with the crust behavior controlling the core, converging to what is effectively a single-component model. This behavior is not seen for $N_t=4630$ in the bottom plot.}
    \label{fig:nc_tauc_ns_taus}
\end{figure}

Interestingly, the output of the expectation-maximization algorithm shows bimodality for $N_t=1157$.
In the top left panel of Fig.~\ref{fig:em_only_measurements:corner_plot} there is a sharp peak near zero in the blue curve for $\Delta t/\tauc$ and a corresponding peak near the right-edge of the allowable region in $\Delta t/\taus$ (seen in the top panel of the second column). These peaks are not evident for $N_t=4630$ (orange curve).
To investigate this bimodality for $N_t=1157$ we show a scatter plot of recovered values for $\Nc/\Ic$ versus $\Delta t/\tauc$ in Fig.~\ref{fig:nc_tauc_ns_taus}. The top and bottom panels correspond to $N_t=1157$ and $N_t=4630$ respectively. It is clear for $N_t=1157$ that there is a cluster of points with $\Delta t/\tauc \approx 10^{-3}$ (which is the low end of the allowable range in Table~\ref{tab:starting_points}), and $\Nc/\Ic \approx \langle\dot\Omega_{\rm c,s}\rangle$ (indicated by the red line), i.e. $\langle\dot\Omega_{\rm c, s}\rangle$ is controlled by the secular torque applied to the crust. The maximum-likelihood estimator converges on a region of parameter space that is closer to a single-component model, where the dynamics of the core follow the dynamics of the crust. 

The cluster of points with $\Nc/\Ic \approx \langle\dot\Omega_{\rm c,s}\rangle$ is also related to the spike at low values of $\tau$ shown in Fig.~\ref{fig:omega_dot_em_only}, which in turn causes the large upper bound on $\sigma_{\rm s}/\Is$ in the fifth column of Table~\ref{tab:estimated_params}. When the inferred value of $\tau$ gets smaller, the inferred amplitude of the noise torques gets larger, because the system can accommodate larger fluctuations in $\Omega_{\rm c}$ and $\Omega_{\rm s}$, which are damped on the observational time-scale $\Delta t$. This is discussed in Appendix~\ref{sec:timing_noise_comparison}, and shown in Fig.~\ref{fig:app:noise_study}. When the peak at low values of $\tau$ goes away, the large upper bound on the estimates of $\sigma_{\rm s}/\Is$ also goes away, as expected.
Finally, if the last terms of (\ref{EQ:CRUST_EQ_OF_MOTION}) and (\ref{EQ:SUPERFLUID_EQ_OF_MOTION}) form an action reaction pair $\taus / \tauc=\Is/\Ic$, the cluster of points in Fig.~\ref{fig:nc_tauc_ns_taus} has $\Is/\Ic\approx 3\times 10^{-3}$. This value is marginally consistent with theoretical expectations, which generally predict $10^{-2} \lesssim \Ic/\Is \lesssim 10^2$~\citep{link:1999iek,chamel:2012cwl}.

The bimodality seen in the parameter estimates is likely caused by the existence of two local maxima in the posterior probability density. The method we have employed does not trace out the shape of the posterior density. It attempts to find the global maximum, sacrificing information for speed. A method that properly samples from the posterior distribution, $p(\bs\theta|\bs y)$, would likely show two peaks of similar height. Directly sampling from the posterior is reserved for future work. We note that direct sampling is more expensive computationally, and may not be suitable for applications involving large numbers of frequently sampled pulsars~\citep{johnston:2020ask}

\section{Conclusion}
\label{sec:conclusion}
In this paper we demonstrate a new way to estimate the parameters of the classic, two-component, crust-superfluid model of a differentially rotating neutron star. We track the timing noise explicitly via  a state-space based identification scheme, instead of subtracting its ensemble statistics. That is, we separate statistically the specific, time-ordered sequence of fluctuations realized in the data into spin wandering intrinsic to the stellar crust (achromatic timing noise) and measurement noise (including chromatic timing noise due to propagation effects) by maximizing the expectation values of the parameters in the noise-driven dynamical model given by~(\ref{EQ:CRUST_EQ_OF_MOTION}) and (\ref{EQ:SUPERFLUID_EQ_OF_MOTION}). In contrast, previous approaches assume that the data are ergodic\footnote{Here ergodic means that the time- and ensemble-averages of a random variable are equal}, use the time-series data to construct the power spectral density of the fluctuations (assumed stationary), and then estimate the parameters of the power spectral density assuming some empirically motivated functional form.
If continuous gravitational waves and electromagnetic radiation trace the angular velocity of different parts of the star, then multi-messenger data plays a key role in our ability to resolve a full picture of the stellar interior. 

Using synthetic data to represent the hypothetical future scenario of simultaneous gravitational-wave measurements of $\omgs$ and electromagnetic measurements of $\omgc$, we accurately estimate the physical parameters $\tauc$, $\taus$, $\Nc/\Ic$, $\Ns/\Is$, $\sigma_{\rm c}/\Ic$, and $\sigma_{\rm s}/\Is$ using our maximum-likelihood method.

When we use only the synthetic electromagnetic measurements of $\omgc$, we accurately estimate two combinations of the original parameters, $\langle\dot\Omega_{\rm c,s}\rangle$, and $\tau$ (defined in [\ref{eq:deterministic_spindown}] and [\ref{eq:tau}] respectively), as well as $\sigma_{\rm c}/\Ic$, for $N_t=4630$. For $N_t=1157$, however, the estimates of $\Delta t/\tauc$, $\Delta t/\taus$ are bimodal. One peak is centered around the correct parameters, while the other peak corresponds to what is effectively a single component model where the crust controls the behavior of the system. This bimodal structure, which disappears with a sufficient amount of data, is likely due to the existence of two local maxima in the posterior probability density. We will properly estimate the full posterior probability density in future work, using methods like Markov chain Monte Carlo.

The method needs some minor modifications before being used on real data. Measurement noise in pulsar-timing analyses is usually considered white in the TOAs, whereas here we treat it as white in the frequency. Extending the model to account for white noise in the frequency and phase would also be useful. Ideally, one would also include a parameter to encapsulate the memory (i.e. ``redness'') of the noise process and estimate it.
Other adjustments to the model, e.g. secular torques which are not constant and depend on $\omgc$ and $\omgs$ are easy to include. The user is encouraged to add or subtract model features as desired.

We present a simple, fast method of parameter estimation using the expectation-maximization algorithm. This method returns the peaks of the posterior distribution. Markov chain Monte Carlo methods like Gibbs sampling and Metropolis-in-Gibbs sampling can be used to estimate the full posterior distribution of the parameters in hidden Markov models~\citep{carlin:1992yud}. Such methods are often complicated by the need to simulate the state variables using the output of the Kalman-filter~\citep{fruhwirth:1994ios,dejong:1995yex,durbin:2002uyt}. We plan to investigate such methods in future work. These methods are also more expensive computationally and may not be suitable for large numbers of pulsars which are monitored frequently, e.g. daily, as in the proposed Thousand Pulsar Array~\citep{johnston:2020ask}, and similar projects~\citep{Jankowski:2019pqd,chime:2020zkd}.

Implementation-related improvements aside, the method of treating timing noise on the same level of the fit parameters is both new and useful. The expectation-maximization method we have presented is general. As long as we can write down a dynamical model in the form of (\ref{eq:inference:state_transition_equation}) and (\ref{eq:inference:measurement_from_states}) we can use this method to quickly estimate parameters. A better understanding of timing noise could lead to improved modeling of neutron stars and increased detection prospects for gravitational-wave searches from pulsar timing arrays~\citep{Hobbs:2017oam}.

\section*{Acknowledgements}
\label{sec:acknowledgements}
The authors acknowledge useful discussions with Sofia Suvorova, who recommended the \texttt{ssm} Matlab package, Robin Evans and William Moran for their help understanding the expectation-maximization algorithm, and Liam Dunn for discussions on integrating the equations of motion. The authors would also like to thank the anonymous referee. Parts of this research were conducted by the Australian Research Council Centre of Excellence for Gravitational Wave Discovery (OzGrav), through project number CE170100004.

\section*{Data Availability}
No new data were generated or analysed in support of this research.

\bibliographystyle{mnras}
\bibliography{references_ads,references_non_ads}

\begin{thebibliography}{}
\makeatletter
\relax
\def\mn@urlcharsother{\let\do\@makeother \do\$\do\&\do\#\do\^\do\_\do\%\do\~}
\def\mn@doi{\begingroup\mn@urlcharsother \@ifnextchar [ {\mn@doi@}
  {\mn@doi@[]}}
\def\mn@doi@[#1]#2{\def\@tempa{#1}\ifx\@tempa\@empty \href
  {http://dx.doi.org/#2} {doi:#2}\else \href {http://dx.doi.org/#2} {#1}\fi
  \endgroup}
\def\mn@eprint#1#2{\mn@eprint@#1:#2::\@nil}
\def\mn@eprint@arXiv#1{\href {http://arxiv.org/abs/#1} {{\tt arXiv:#1}}}
\def\mn@eprint@dblp#1{\href {http://dblp.uni-trier.de/rec/bibtex/#1.xml}
  {dblp:#1}}
\def\mn@eprint@#1:#2:#3:#4\@nil{\def\@tempa {#1}\def\@tempb {#2}\def\@tempc
  {#3}\ifx \@tempc \@empty \let \@tempc \@tempb \let \@tempb \@tempa \fi \ifx
  \@tempb \@empty \def\@tempb {arXiv}\fi \@ifundefined
  {mn@eprint@\@tempb}{\@tempb:\@tempc}{\expandafter \expandafter \csname
  mn@eprint@\@tempb\endcsname \expandafter{\@tempc}}}

\bibitem[\protect\citeauthoryear{Aasi et~al.}{Aasi
  et~al.}{2015}]{TheLIGOScientific:2014jea}
Aasi J.,  et~al., 2015, \mn@doi [Class. Quant. Grav.]
  {10.1088/0264-9381/32/7/074001}, 32, 074001

\bibitem[\protect\citeauthoryear{{Abbott} et~al.}{{Abbott}
  et~al.}{2017a}]{2017PhRvL.119p1101A}
{Abbott} B.~P.,  et~al., 2017a, \mn@doi [\prl]
  {10.1103/PhysRevLett.119.161101}, \href
  {https://ui.adsabs.harvard.edu/abs/2017PhRvL.119p1101A} {119, 161101}

\bibitem[\protect\citeauthoryear{{Abbott} et~al.}{{Abbott}
  et~al.}{2017b}]{2017ApJ...848L..12A}
{Abbott} B.~P.,  et~al., 2017b, \mn@doi [\apjl] {10.3847/2041-8213/aa91c9},
  \href {https://ui.adsabs.harvard.edu/abs/2017ApJ...848L..12A} {848, L12}

\bibitem[\protect\citeauthoryear{{Abbott} et~al.}{{Abbott}
  et~al.}{2018}]{2018PhRvL.121p1101A}
{Abbott} B.~P.,  et~al., 2018, \mn@doi [\prl] {10.1103/PhysRevLett.121.161101},
  \href {https://ui.adsabs.harvard.edu/abs/2018PhRvL.121p1101A} {121, 161101}

\bibitem[\protect\citeauthoryear{{Abney} \& {Epstein}}{{Abney} \&
  {Epstein}}{1996}]{Abney:1996aaaa}
{Abney} M.,  {Epstein} R.~I.,  1996, \mn@doi [Journal of Fluid Mechanics]
  {10.1017/S0022112096002030}, \href
  {https://ui.adsabs.harvard.edu/abs/1996JFM...312..327A} {312, 327}

\bibitem[\protect\citeauthoryear{Acernese et~al.}{Acernese
  et~al.}{2015}]{TheVirgo:2014hva}
Acernese F.,  et~al., 2015, \mn@doi [Class. Quant. Grav.]
  {10.1088/0264-9381/32/2/024001}, 32, 024001

\bibitem[\protect\citeauthoryear{Alford}{Alford}{2001}]{Alford2001}
Alford M.,  2001, \mn@doi [Annual Review of Nuclear and Particle Science]
  {10.1146/annurev.nucl.51.101701.132449}, 51, 131

\bibitem[\protect\citeauthoryear{{Alpar}, {Pines}, {Anderson}  \&
  {Shaham}}{{Alpar} et~al.}{1984a}]{alpar:1984yui}
{Alpar} M.~A.,  {Pines} D.,  {Anderson} P.~W.,   {Shaham} J.,  1984a, \mn@doi
  [\apj] {10.1086/161616}, \href
  {https://ui.adsabs.harvard.edu/abs/1984ApJ...276..325A} {276, 325}

\bibitem[\protect\citeauthoryear{{Alpar}, {Anderson}, {Pines}  \&
  {Shaham}}{{Alpar} et~al.}{1984b}]{alpar:1984hrw}
{Alpar} M.~A.,  {Anderson} P.~W.,  {Pines} D.,   {Shaham} J.,  1984b, \mn@doi
  [\apj] {10.1086/161849}, \href
  {https://ui.adsabs.harvard.edu/abs/1984ApJ...278..791A} {278, 791}

\bibitem[\protect\citeauthoryear{{Alpar}, {Nandkumar}  \& {Pines}}{{Alpar}
  et~al.}{1986}]{1986ApJ...311..197A}
{Alpar} M.~A.,  {Nandkumar} R.,   {Pines} D.,  1986, \mn@doi [\apj]
  {10.1086/164765}, \href
  {https://ui.adsabs.harvard.edu/abs/1986ApJ...311..197A} {311, 197}

\bibitem[\protect\citeauthoryear{{Alpar}, {Chau}, {Cheng}  \& {Pines}}{{Alpar}
  et~al.}{1996}]{alpar:1996aew}
{Alpar} M.~A.,  {Chau} H.~F.,  {Cheng} K.~S.,   {Pines} D.,  1996, \mn@doi
  [\apj] {10.1086/176935}, \href
  {https://ui.adsabs.harvard.edu/abs/1996ApJ...459..706A} {459, 706}

\bibitem[\protect\citeauthoryear{{Andersson} \& {Comer}}{{Andersson} \&
  {Comer}}{2006}]{2006CQGra..23.5505A}
{Andersson} N.,  {Comer} G.~L.,  2006, \mn@doi [Classical and Quantum Gravity]
  {10.1088/0264-9381/23/18/003}, \href
  {https://ui.adsabs.harvard.edu/abs/2006CQGra..23.5505A} {23, 5505}

\bibitem[\protect\citeauthoryear{{Andersson}, {Glampedakis}, {Ho}  \&
  {Espinoza}}{{Andersson} et~al.}{2012}]{2012PhRvL.109x1103A}
{Andersson} N.,  {Glampedakis} K.,  {Ho} W.~C.~G.,   {Espinoza} C.~M.,  2012,
  \mn@doi [\prl] {10.1103/PhysRevLett.109.241103}, \href
  {https://ui.adsabs.harvard.edu/abs/2012PhRvL.109x1103A} {109, 241103}

\bibitem[\protect\citeauthoryear{{Andreev} \& {Bashkin}}{{Andreev} \&
  {Bashkin}}{1976}]{1976JETP...42..164A}
{Andreev} A.~F.,  {Bashkin} E.~P.,  1976, Soviet Journal of Experimental and
  Theoretical Physics, \href
  {https://ui.adsabs.harvard.edu/abs/1976JETP...42..164A} {42, 164}

\bibitem[\protect\citeauthoryear{{Antonelli} \& {Pizzochero}}{{Antonelli} \&
  {Pizzochero}}{2017}]{Antonelli:2017oxd}
{Antonelli} M.,  {Pizzochero} P.~M.,  2017, \mn@doi [\mnras]
  {10.1093/mnras/stw2376}, \href
  {https://ui.adsabs.harvard.edu/abs/2017MNRAS.464..721A} {464, 721}

\bibitem[\protect\citeauthoryear{{Anzuini} \& {Melatos}}{{Anzuini} \&
  {Melatos}}{2020}]{2020MNRAS.494.3095A}
{Anzuini} F.,  {Melatos} A.,  2020, \mn@doi [\mnras] {10.1093/mnras/staa915},
  \href {https://ui.adsabs.harvard.edu/abs/2020MNRAS.494.3095A} {494, 3095}

\bibitem[\protect\citeauthoryear{{Archibald}, {Kondratiev}, {Hessels}  \&
  {Stinebring}}{{Archibald} et~al.}{2014}]{2014ApJ...790L..22A}
{Archibald} A.~M.,  {Kondratiev} V.~I.,  {Hessels} J. W.~T.,   {Stinebring}
  D.~R.,  2014, \mn@doi [\apjl] {10.1088/2041-8205/790/2/L22}, \href
  {https://ui.adsabs.harvard.edu/abs/2014ApJ...790L..22A} {790, L22}

\bibitem[\protect\citeauthoryear{{Arzoumanian}, {Nice}, {Taylor}  \&
  {Thorsett}}{{Arzoumanian} et~al.}{1994}]{1994ApJ...422..671A}
{Arzoumanian} Z.,  {Nice} D.~J.,  {Taylor} J.~H.,   {Thorsett} S.~E.,  1994,
  \mn@doi [\apj] {10.1086/173760}, \href
  {https://ui.adsabs.harvard.edu/abs/1994ApJ...422..671A} {422, 671}

\bibitem[\protect\citeauthoryear{{Ashton}, {Lasky}, {Graber}  \&
  {Palfreyman}}{{Ashton} et~al.}{2019}]{2019NatAs...3.1143A}
{Ashton} G.,  {Lasky} P.~D.,  {Graber} V.,   {Palfreyman} J.,  2019, \mn@doi
  [Nature Astronomy] {10.1038/s41550-019-0844-6}, \href
  {https://ui.adsabs.harvard.edu/abs/2019NatAs...3.1143A} {3, 1143}

\bibitem[\protect\citeauthoryear{Aso, Michimura, Somiya, Ando, Miyakawa,
  Sekiguchi, Tatsumi  \& Yamamoto}{Aso et~al.}{2013}]{PhysRevD.88.043007}
Aso Y.,  Michimura Y.,  Somiya K.,  Ando M.,  Miyakawa O.,  Sekiguchi T.,
  Tatsumi D.,   Yamamoto H.,  2013, \mn@doi [Phys. Rev. D]
  {10.1103/PhysRevD.88.043007}, 88, 043007

\bibitem[\protect\citeauthoryear{{Avogadro}, {Barranco}, {Broglia}  \&
  {Vigezzi}}{{Avogadro} et~al.}{2008}]{Avogadro:2008poi}
{Avogadro} P.,  {Barranco} F.,  {Broglia} R.~A.,   {Vigezzi} E.,  2008, \mn@doi
  [\nphysa] {10.1016/j.nuclphysa.2008.07.010}, \href
  {https://ui.adsabs.harvard.edu/abs/2008NuPhA.811..378A} {811, 378}

\bibitem[\protect\citeauthoryear{{Barsukov}, {Goglichidze}  \&
  {Tsygan}}{{Barsukov} et~al.}{2014}]{2014MNRAS.444.1318B}
{Barsukov} D.~P.,  {Goglichidze} O.~A.,   {Tsygan} A.~I.,  2014, \mn@doi
  [\mnras] {10.1093/mnras/stu1516}, \href
  {https://ui.adsabs.harvard.edu/abs/2014MNRAS.444.1318B} {444, 1318}

\bibitem[\protect\citeauthoryear{{Baykal}}{{Baykal}}{1997}]{baykal:1997lvw}
{Baykal} A.,  1997, \aap, \href
  {https://ui.adsabs.harvard.edu/abs/1997A&A...319..515B} {319, 515}

\bibitem[\protect\citeauthoryear{{Baykal} \& {Oegelman}}{{Baykal} \&
  {Oegelman}}{1993}]{Baykal:1993aaa}
{Baykal} A.,  {Oegelman} H.,  1993, \aap, \href
  {https://ui.adsabs.harvard.edu/abs/1993A&A...267..119B} {267, 119}

\bibitem[\protect\citeauthoryear{{Baykal}, {Alpar}  \& {Kiziloglu}}{{Baykal}
  et~al.}{1991}]{Baykal:1991uec}
{Baykal} A.,  {Alpar} A.,   {Kiziloglu} U.,  1991, \aap, \href
  {https://ui.adsabs.harvard.edu/abs/1991A&A...252..664B} {252, 664}

\bibitem[\protect\citeauthoryear{{Baym}, {Pethick}  \& {Pines}}{{Baym}
  et~al.}{1969a}]{1969Natur.224..673B}
{Baym} G.,  {Pethick} C.,   {Pines} D.,  1969a, \mn@doi [\nat]
  {10.1038/224673a0}, \href
  {https://ui.adsabs.harvard.edu/abs/1969Natur.224..673B} {224, 673}

\bibitem[\protect\citeauthoryear{{Baym}, {Pethick}, {Pines}  \&
  {Ruderman}}{{Baym} et~al.}{1969b}]{1969Natur.224..872B}
{Baym} G.,  {Pethick} C.,  {Pines} D.,   {Ruderman} M.,  1969b, \mn@doi [\nat]
  {10.1038/224872a0}, \href
  {https://ui.adsabs.harvard.edu/abs/1969Natur.224..872B} {224, 872}

\bibitem[\protect\citeauthoryear{{Bhattacharya} \& {Srinivasan}}{{Bhattacharya}
  \& {Srinivasan}}{1991}]{Bhattacharya:1991lsds}
{Bhattacharya} D.,  {Srinivasan} G.,  1991, in {Ventura} J.,  {Pines} D.,  eds,
   NATO Advanced Science Institutes (ASI) Series C Vol. 344, NATO Advanced
  Science Institutes (ASI) Series C. p.~219

\bibitem[\protect\citeauthoryear{{Bildsten} et~al.,}{{Bildsten}
  et~al.}{1997}]{1997ApJS..113..367B}
{Bildsten} L.,  et~al., 1997, \mn@doi [\apjs] {10.1086/313060}, \href
  {https://ui.adsabs.harvard.edu/abs/1997ApJS..113..367B} {113, 367}

\bibitem[\protect\citeauthoryear{Bogdanov et~al.,}{Bogdanov
  et~al.}{2019}]{Bogdanov_2019}
Bogdanov S.,  et~al., 2019, \mn@doi [The Astrophysical Journal]
  {10.3847/2041-8213/ab53eb}, 887, L25

\bibitem[\protect\citeauthoryear{{CHIME/Pulsar Collaboration}
  et~al.,}{{CHIME/Pulsar Collaboration} et~al.}{2020}]{chime:2020zkd}
{CHIME/Pulsar Collaboration} et~al., 2020, arXiv e-prints, \href
  {https://ui.adsabs.harvard.edu/abs/2020arXiv200805681C} {p. arXiv:2008.05681}

\bibitem[\protect\citeauthoryear{Carlin, Polson  \& Stoffer}{Carlin
  et~al.}{1992}]{carlin:1992yud}
Carlin B.~P.,  Polson N.~G.,   Stoffer D.~S.,  1992, Journal of the American
  Statistical Association, 87, 493

\bibitem[\protect\citeauthoryear{{\c C}erri-Serim, Serim, {\c S}ahiner, {\.
  I}nam  \& Baykal}{{\c C}erri-Serim et~al.}{2019}]{10.1093/mnras/sty3213}
{\c C}erri-Serim D.,  Serim M.~M.,  {\c S}ahiner {\c S}.,  {\. I}nam S.~{\c
  C}.,   Baykal A.,  2019, \mn@doi [\mnras] {10.1093/mnras/sty3213}, 485, 2

\bibitem[\protect\citeauthoryear{{Chamel}}{{Chamel}}{2012}]{chamel:2012cwl}
{Chamel} N.,  2012, \mn@doi [\prc] {10.1103/PhysRevC.85.035801}, \href
  {https://ui.adsabs.harvard.edu/abs/2012PhRvC..85c5801C} {85, 035801}

\bibitem[\protect\citeauthoryear{{Chamel}}{{Chamel}}{2013}]{2013PhRvL.110a1101C}
{Chamel} N.,  2013, \mn@doi [\prl] {10.1103/PhysRevLett.110.011101}, \href
  {https://ui.adsabs.harvard.edu/abs/2013PhRvL.110a1101C} {110, 011101}

\bibitem[\protect\citeauthoryear{{Chamel}}{{Chamel}}{2017}]{2017JApA...38...43C}
{Chamel} N.,  2017, \mn@doi [Journal of Astrophysics and Astronomy]
  {10.1007/s12036-017-9470-9}, \href
  {https://ui.adsabs.harvard.edu/abs/2017JApA...38...43C} {38, 43}

\bibitem[\protect\citeauthoryear{{Cheng}}{{Cheng}}{1987a}]{cheng:1987ekv}
{Cheng} K.~S.,  1987a, \mn@doi [\apj] {10.1086/165672}, \href
  {https://ui.adsabs.harvard.edu/abs/1987ApJ...321..799C} {321, 799}

\bibitem[\protect\citeauthoryear{{Cheng}}{{Cheng}}{1987b}]{cheng:1987uid}
{Cheng} K.~S.,  1987b, \mn@doi [\apj] {10.1086/165673}, \href
  {https://ui.adsabs.harvard.edu/abs/1987ApJ...321..805C} {321, 805}

\bibitem[\protect\citeauthoryear{{Cordes}}{{Cordes}}{1980}]{1980ApJ...237..216C}
{Cordes} J.~M.,  1980, \mn@doi [\apj] {10.1086/157861}, \href
  {https://ui.adsabs.harvard.edu/abs/1980ApJ...237..216C} {237, 216}

\bibitem[\protect\citeauthoryear{{Cordes}}{{Cordes}}{2013}]{2013CQGra..30v4002C}
{Cordes} J.~M.,  2013, \mn@doi [Classical and Quantum Gravity]
  {10.1088/0264-9381/30/22/224002}, \href
  {https://ui.adsabs.harvard.edu/abs/2013CQGra..30v4002C} {30, 224002}

\bibitem[\protect\citeauthoryear{{Cordes} \& {Helfand}}{{Cordes} \&
  {Helfand}}{1980}]{1980ApJ...239..640C}
{Cordes} J.~M.,  {Helfand} D.~J.,  1980, \mn@doi [\apj] {10.1086/158150}, \href
  {https://ui.adsabs.harvard.edu/abs/1980ApJ...239..640C} {239, 640}

\bibitem[\protect\citeauthoryear{{Coughlin}, {Dietrich}, {Margalit}  \&
  {Metzger}}{{Coughlin} et~al.}{2019}]{2019MNRAS.489L..91C}
{Coughlin} M.~W.,  {Dietrich} T.,  {Margalit} B.,   {Metzger} B.~D.,  2019,
  \mn@doi [\mnras] {10.1093/mnrasl/slz133}, \href
  {https://ui.adsabs.harvard.edu/abs/2019MNRAS.489L..91C} {489, L91}

\bibitem[\protect\citeauthoryear{{D'Angelo} \& {Spruit}}{{D'Angelo} \&
  {Spruit}}{2010}]{dangelo:2010esf}
{D'Angelo} C.~R.,  {Spruit} H.~C.,  2010, \mn@doi [\mnras]
  {10.1111/j.1365-2966.2010.16749.x}, \href
  {https://ui.adsabs.harvard.edu/abs/2010MNRAS.406.1208D} {406, 1208}

\bibitem[\protect\citeauthoryear{{Deeter}}{{Deeter}}{1984}]{1984ApJ...281..482D}
{Deeter} J.~E.,  1984, \mn@doi [\apj] {10.1086/162122}, \href
  {https://ui.adsabs.harvard.edu/abs/1984ApJ...281..482D} {281, 482}

\bibitem[\protect\citeauthoryear{Dempster, Laird  \& Rubin}{Dempster
  et~al.}{1977}]{em_original_paper}
Dempster A.~P.,  Laird N.~M.,   Rubin D.~B.,  1977, \mn@doi [Journal of the
  Royal Statistical Society: Series B (Methodological)]
  {10.1111/j.2517-6161.1977.tb01600.x}, 39, 1

\bibitem[\protect\citeauthoryear{{Digalakis}, {Rohlicek}  \&
  {Ostendorf}}{{Digalakis} et~al.}{1993}]{Digalakis_1993_EM_algorithm}
{Digalakis} V.,  {Rohlicek} J.~R.,   {Ostendorf} M.,  1993, \mn@doi [IEEE
  Transactions on Speech and Audio Processing] {10.1109/89.242489}, 1, 431

\bibitem[\protect\citeauthoryear{{Dolch} et~al.,}{{Dolch}
  et~al.}{2020}]{2020arXiv200810562D}
{Dolch} T.,  et~al., 2020, arXiv e-prints, \href
  {https://ui.adsabs.harvard.edu/abs/2020arXiv200810562D} {p. arXiv:2008.10562}

\bibitem[\protect\citeauthoryear{{Drummond} \& {Melatos}}{{Drummond} \&
  {Melatos}}{2017}]{drummond:2017wfg}
{Drummond} L.~V.,  {Melatos} A.,  2017, \mn@doi [\mnras]
  {10.1093/mnras/stx2301}, \href
  {https://ui.adsabs.harvard.edu/abs/2017MNRAS.472.4851D} {472, 4851}

\bibitem[\protect\citeauthoryear{{Drummond} \& {Melatos}}{{Drummond} \&
  {Melatos}}{2018}]{drummond:2018esf}
{Drummond} L.~V.,  {Melatos} A.,  2018, \mn@doi [\mnras]
  {10.1093/mnras/stx3197}, \href
  {https://ui.adsabs.harvard.edu/abs/2018MNRAS.475..910D} {475, 910}

\bibitem[\protect\citeauthoryear{Durbin \& Koopman}{Durbin \&
  Koopman}{2002}]{durbin:2002uyt}
Durbin J.,  Koopman S.~J.,  2002, Biometrika, 89, 603

\bibitem[\protect\citeauthoryear{Durbin \& Koopman}{Durbin \&
  Koopman}{2012}]{durbin2012time}
Durbin J.,  Koopman S.,  2012, Time Series Analysis by State Space Methods:
  Second Edition.
Oxford Statistical Science Series, OUP Oxford, \url
  {https://books.google.com.au/books?id=fOq39Zh0olQC}

\bibitem[\protect\citeauthoryear{{Easson}}{{Easson}}{1979}]{Easson:1979edf}
{Easson} I.,  1979, \mn@doi [\apj] {10.1086/157432}, \href
  {https://ui.adsabs.harvard.edu/abs/1979ApJ...233..711E} {233, 711}

\bibitem[\protect\citeauthoryear{{Espinoza}, {Lyne}, {Stappers}  \&
  {Kramer}}{{Espinoza} et~al.}{2011}]{2011MNRAS.414.1679E}
{Espinoza} C.~M.,  {Lyne} A.~G.,  {Stappers} B.~W.,   {Kramer} M.,  2011,
  \mn@doi [\mnras] {10.1111/j.1365-2966.2011.18503.x}, \href
  {https://ui.adsabs.harvard.edu/abs/2011MNRAS.414.1679E} {414, 1679}

\bibitem[\protect\citeauthoryear{Frühwirth-Schnatter}{Frühwirth-Schnatter}{1994}]{fruhwirth:1994ios}
Frühwirth-Schnatter S.,  1994, \mn@doi [Journal of Time Series Analysis]
  {10.1111/j.1467-9892.1994.tb00184.x}, 15, 183

\bibitem[\protect\citeauthoryear{{Fulgenzi}, {Melatos}  \& {Hughes}}{{Fulgenzi}
  et~al.}{2017}]{fulgenzi:2017lpw}
{Fulgenzi} W.,  {Melatos} A.,   {Hughes} B.~D.,  2017, \mn@doi [\mnras]
  {10.1093/mnras/stx1353}, \href
  {https://ui.adsabs.harvard.edu/abs/2017MNRAS.470.4307F} {470, 4307}

\bibitem[\protect\citeauthoryear{Gelman, Carlin, Stern, Dunson, Vehtari  \&
  Rubin}{Gelman et~al.}{2013}]{gelman2013bayesian}
Gelman A.,  Carlin J.,  Stern H.,  Dunson D.,  Vehtari A.,   Rubin D.,  2013,
  Bayesian Data Analysis, Third Edition.
Chapman \& Hall/CRC Texts in Statistical Science, Taylor \& Francis, \url
  {https://books.google.com.au/books?id=ZXL6AQAAQBAJ}

\bibitem[\protect\citeauthoryear{{Gendreau}, {Arzoumanian}  et~al.}{{Gendreau}
  et~al.}{2016}]{10.1117/12.2231304}
{Gendreau} K.~C.,  {Arzoumanian} Z.,   et~al., 2016, in {den Herder} J.-W.~A.,
  {Takahashi} T.,   {Bautz} M.,  eds,  Society of Photo-Optical Instrumentation
  Engineers (SPIE) Conference Series Vol. 9905, Space Telescopes and
  Instrumentation 2016: Ultraviolet to Gamma Ray. p. 99051H,
  \mn@doi{10.1117/12.2231304}

\bibitem[\protect\citeauthoryear{Ghosh \& Lamb}{Ghosh \&
  Lamb}{1979}]{Ghosh1979}
Ghosh P.,  Lamb F.~K.,  1979, \mn@doi [\apj] {10.1086/157498}, 234, 296

\bibitem[\protect\citeauthoryear{Gibson \& Ninness}{Gibson \&
  Ninness}{2005}]{GIBSON20051667}
Gibson S.,  Ninness B.,  2005, \mn@doi [Automatica]
  {https://doi.org/10.1016/j.automatica.2005.05.008}, 41, 1667

\bibitem[\protect\citeauthoryear{{Glampedakis} \& {Lasky}}{{Glampedakis} \&
  {Lasky}}{2015}]{glampedakis:2015yhf}
{Glampedakis} K.,  {Lasky} P.~D.,  2015, \mn@doi [\mnras]
  {10.1093/mnras/stv638}, \href
  {https://ui.adsabs.harvard.edu/abs/2015MNRAS.450.1638G} {450, 1638}

\bibitem[\protect\citeauthoryear{{Glampedakis}, {Andersson}  \&
  {Samuelsson}}{{Glampedakis} et~al.}{2011}]{2011MNRAS.410..805G}
{Glampedakis} K.,  {Andersson} N.,   {Samuelsson} L.,  2011, \mn@doi [\mnras]
  {10.1111/j.1365-2966.2010.17484.x}, \href
  {https://ui.adsabs.harvard.edu/abs/2011MNRAS.410..805G} {410, 805}

\bibitem[\protect\citeauthoryear{Goglichidze \& Barsukov}{Goglichidze \&
  Barsukov}{2018}]{10.1093/mnras/sty2864}
Goglichidze O.~A.,  Barsukov D.~P.,  2018, \mn@doi [Monthly Notices of the
  Royal Astronomical Society] {10.1093/mnras/sty2864}, 482, 3032

\bibitem[\protect\citeauthoryear{{Goldreich} \& {Julian}}{{Goldreich} \&
  {Julian}}{1969}]{goldreich:1969wiu}
{Goldreich} P.,  {Julian} W.~H.,  1969, \mn@doi [\apj] {10.1086/150119}, \href
  {https://ui.adsabs.harvard.edu/abs/1969ApJ...157..869G} {157, 869}

\bibitem[\protect\citeauthoryear{{Goncharov}, {Zhu}  \& {Thrane}}{{Goncharov}
  et~al.}{2019}]{2019arXiv191005961G}
{Goncharov} B.,  {Zhu} X.-J.,   {Thrane} E.,  2019, arXiv e-prints, \href
  {https://ui.adsabs.harvard.edu/abs/2019arXiv191005961G} {p. arXiv:1910.05961}

\bibitem[\protect\citeauthoryear{{Goncharov} et~al.,}{{Goncharov}
  et~al.}{2020}]{2020MNRAS.tmp.3250G}
{Goncharov} B.,  et~al., 2020, \mn@doi [\mnras] {10.1093/mnras/staa3411}, \href
  {https://ui.adsabs.harvard.edu/abs/2020MNRAS.tmp.3250G} {}

\bibitem[\protect\citeauthoryear{Gorter \& Mellink}{Gorter \&
  Mellink}{1949}]{GORTER1949285}
Gorter C.,  Mellink J.,  1949, \mn@doi [Physica]
  {https://doi.org/10.1016/0031-8914(49)90105-6}, 15, 285

\bibitem[\protect\citeauthoryear{{Graber}, {Andersson}  \& {Hogg}}{{Graber}
  et~al.}{2017}]{2017IJMPD..2630015G}
{Graber} V.,  {Andersson} N.,   {Hogg} M.,  2017, \mn@doi [International
  Journal of Modern Physics D] {10.1142/S0218271817300154}, \href
  {https://ui.adsabs.harvard.edu/abs/2017IJMPD..2630015G} {26, 1730015}

\bibitem[\protect\citeauthoryear{{Graber}, {Cumming}  \& {Andersson}}{{Graber}
  et~al.}{2018}]{2018ApJ...865...23G}
{Graber} V.,  {Cumming} A.,   {Andersson} N.,  2018, \mn@doi [\apj]
  {10.3847/1538-4357/aad776}, \href
  {https://ui.adsabs.harvard.edu/abs/2018ApJ...865...23G} {865, 23}

\bibitem[\protect\citeauthoryear{Greenstein}{Greenstein}{1970}]{GREENSTEIN_1970}
Greenstein G.,  1970, \mn@doi [Nature]
  {https://ui.adsabs.harvard.edu/link_gateway/1970Natur.227..791G/doi:10.1038/227791a0},
  227, 791

\bibitem[\protect\citeauthoryear{{Groth}}{{Groth}}{1975}]{1975ApJS...29..453G}
{Groth} E.~J.,  1975, \mn@doi [\apjs] {10.1086/190354}, \href
  {https://ui.adsabs.harvard.edu/abs/1975ApJS...29..453G} {29, 453}

\bibitem[\protect\citeauthoryear{{G{\"u}gercino{\u{g}}lu} \&
  {Alpar}}{{G{\"u}gercino{\u{g}}lu} \& {Alpar}}{2014}]{Gugercinoglu:2014als}
{G{\"u}gercino{\u{g}}lu} E.,  {Alpar} M.~A.,  2014, \mn@doi [\apjl]
  {10.1088/2041-8205/788/1/L11}, \href
  {https://ui.adsabs.harvard.edu/abs/2014ApJ...788L..11G} {788, L11}

\bibitem[\protect\citeauthoryear{{G{\"u}gercino{\v{g}}lu} \&
  {Alpar}}{{G{\"u}gercino{\v{g}}lu} \& {Alpar}}{2017}]{2017MNRAS.471.4827G}
{G{\"u}gercino{\v{g}}lu} E.,  {Alpar} M.~A.,  2017, \mn@doi [\mnras]
  {10.1093/mnras/stx1937}, \href
  {https://ui.adsabs.harvard.edu/abs/2017MNRAS.471.4827G} {471, 4827}

\bibitem[\protect\citeauthoryear{{Haskell} \& {Melatos}}{{Haskell} \&
  {Melatos}}{2015}]{2015IJMPD..2430008H}
{Haskell} B.,  {Melatos} A.,  2015, \mn@doi [International Journal of Modern
  Physics D] {10.1142/S0218271815300086}, \href
  {https://ui.adsabs.harvard.edu/abs/2015IJMPD..2430008H} {24, 1530008}

\bibitem[\protect\citeauthoryear{{Haskell}, {Jones}  \& {Andersson}}{{Haskell}
  et~al.}{2006}]{haskell:2006oiu}
{Haskell} B.,  {Jones} D.~I.,   {Andersson} N.,  2006, \mn@doi [\mnras]
  {10.1111/j.1365-2966.2006.10998.x}, \href
  {https://ui.adsabs.harvard.edu/abs/2006MNRAS.373.1423H} {373, 1423}

\bibitem[\protect\citeauthoryear{Haskell, Pizzochero  \& Sidery}{Haskell
  et~al.}{2012}]{haskell:2012rds}
Haskell B.,  Pizzochero P.~M.,   Sidery T.,  2012, \mn@doi [Monthly Notices of
  the Royal Astronomical Society] {10.1111/j.1365-2966.2011.20080.x}, 420, 658

\bibitem[\protect\citeauthoryear{{Haskell}, {Pizzochero}  \&
  {Seveso}}{{Haskell} et~al.}{2013}]{haskell:2013sdd}
{Haskell} B.,  {Pizzochero} P.~M.,   {Seveso} S.,  2013, \mn@doi [\apjl]
  {10.1088/2041-8205/764/2/L25}, \href
  {https://ui.adsabs.harvard.edu/abs/2013ApJ...764L..25H} {764, L25}

\bibitem[\protect\citeauthoryear{{Ho}, {Wijngaarden}, {Chang}, {Heinke},
  {Page}, {Beznogov}  \& {Patnaude}}{{Ho} et~al.}{2019}]{2019AIPC.2127b0007H}
{Ho} W. C.~G.,  {Wijngaarden} M.~J.~P.,  {Chang} P.,  {Heinke} C.~O.,  {Page}
  D.,  {Beznogov} M.,   {Patnaude} D.~J.,  2019, in American Institute of
  Physics Conference Series. p. 020007 (\mn@eprint {arXiv} {1904.07505}),
  \mn@doi{10.1063/1.5117797}

\bibitem[\protect\citeauthoryear{Hobbs \& Dai}{Hobbs \&
  Dai}{2017}]{Hobbs:2017oam}
Hobbs G.,  Dai S.,  2017, \mn@doi [Natl. Sci. Rev.] {10.1093/nsr/nwx126}, 4,
  707

\bibitem[\protect\citeauthoryear{{Hobbs}, {Lyne}  \& {Kramer}}{{Hobbs}
  et~al.}{2010}]{2010MNRAS.402.1027H}
{Hobbs} G.,  {Lyne} A.~G.,   {Kramer} M.,  2010, \mn@doi [\mnras]
  {10.1111/j.1365-2966.2009.15938.x}, \href
  {https://ui.adsabs.harvard.edu/abs/2010MNRAS.402.1027H} {402, 1027}

\bibitem[\protect\citeauthoryear{{Howitt}, {Haskell}  \& {Melatos}}{{Howitt}
  et~al.}{2016}]{2016MNRAS.460.1201H}
{Howitt} G.,  {Haskell} B.,   {Melatos} A.,  2016, \mn@doi [\mnras]
  {10.1093/mnras/stw1043}, \href
  {https://ui.adsabs.harvard.edu/abs/2016MNRAS.460.1201H} {460, 1201}

\bibitem[\protect\citeauthoryear{{Jankowski} et~al.,}{{Jankowski}
  et~al.}{2019}]{Jankowski:2019pqd}
{Jankowski} F.,  et~al., 2019, \mn@doi [\mnras] {10.1093/mnras/sty3390}, \href
  {https://ui.adsabs.harvard.edu/abs/2019MNRAS.484.3691J} {484, 3691}

\bibitem[\protect\citeauthoryear{{Johnston} et~al.,}{{Johnston}
  et~al.}{2020}]{johnston:2020ask}
{Johnston} S.,  et~al., 2020, \mn@doi [\mnras] {10.1093/mnras/staa516}, \href
  {https://ui.adsabs.harvard.edu/abs/2020MNRAS.493.3608J} {493, 3608}

\bibitem[\protect\citeauthoryear{{Jones}}{{Jones}}{2010}]{2010MNRAS.402.2503J}
{Jones} D.~I.,  2010, \mn@doi [\mnras] {10.1111/j.1365-2966.2009.16059.x},
  \href {https://ui.adsabs.harvard.edu/abs/2010MNRAS.402.2503J} {402, 2503}

\bibitem[\protect\citeauthoryear{Kalman}{Kalman}{1960}]{kalman1960}
Kalman R.~E.,  1960, Transactions of the ASME--Journal of Basic Engineering,
  82, 35

\bibitem[\protect\citeauthoryear{{Keith} et~al.,}{{Keith}
  et~al.}{2013}]{2013MNRAS.429.2161K}
{Keith} M.~J.,  et~al., 2013, \mn@doi [\mnras] {10.1093/mnras/sts486}, \href
  {https://ui.adsabs.harvard.edu/abs/2013MNRAS.429.2161K} {429, 2161}

\bibitem[\protect\citeauthoryear{{Khomenko} \& {Haskell}}{{Khomenko} \&
  {Haskell}}{2018}]{khomenko:2018sdf}
{Khomenko} V.,  {Haskell} B.,  2018, \mn@doi [\pasa] {10.1017/pasa.2018.12},
  \href {https://ui.adsabs.harvard.edu/abs/2018PASA...35...20K} {35, e020}

\bibitem[\protect\citeauthoryear{Kristensen, Madsen  \& Jørgensen}{Kristensen
  et~al.}{2004}]{KRISTENSEN2004225}
Kristensen N.~R.,  Madsen H.,   Jørgensen S.~B.,  2004, \mn@doi [Automatica]
  {https://doi.org/10.1016/j.automatica.2003.10.001}, 40, 225

\bibitem[\protect\citeauthoryear{{Lam} et~al.,}{{Lam}
  et~al.}{2017}]{2017ApJ...834...35L}
{Lam} M.~T.,  et~al., 2017, \mn@doi [\apj] {10.3847/1538-4357/834/1/35}, \href
  {https://ui.adsabs.harvard.edu/abs/2017ApJ...834...35L} {834, 35}

\bibitem[\protect\citeauthoryear{{Lasky}, {Melatos}, {Ravi}  \&
  {Hobbs}}{{Lasky} et~al.}{2015}]{lasky:2015pfd}
{Lasky} P.~D.,  {Melatos} A.,  {Ravi} V.,   {Hobbs} G.,  2015, \mn@doi [\mnras]
  {10.1093/mnras/stv540}, \href
  {https://ui.adsabs.harvard.edu/abs/2015MNRAS.449.3293L} {449, 3293}

\bibitem[\protect\citeauthoryear{{Lattimer} \& {Prakash}}{{Lattimer} \&
  {Prakash}}{2016}]{2016PhR...621..127L}
{Lattimer} J.~M.,  {Prakash} M.,  2016, \mn@doi [\physrep]
  {10.1016/j.physrep.2015.12.005}, \href
  {https://ui.adsabs.harvard.edu/abs/2016PhR...621..127L} {621, 127}

\bibitem[\protect\citeauthoryear{{Lentati} et~al.,}{{Lentati}
  et~al.}{2016}]{2016MNRAS.458.2161L}
{Lentati} L.,  et~al., 2016, \mn@doi [\mnras] {10.1093/mnras/stw395}, \href
  {https://ui.adsabs.harvard.edu/abs/2016MNRAS.458.2161L} {458, 2161}

\bibitem[\protect\citeauthoryear{{Levin} et~al.,}{{Levin}
  et~al.}{2016}]{2016ApJ...818..166L}
{Levin} L.,  et~al., 2016, \mn@doi [\apj] {10.3847/0004-637X/818/2/166}, \href
  {https://ui.adsabs.harvard.edu/abs/2016ApJ...818..166L} {818, 166}

\bibitem[\protect\citeauthoryear{{Link} \& {Epstein}}{{Link} \&
  {Epstein}}{1991}]{Link:1991dlp}
{Link} B.~K.,  {Epstein} R.~I.,  1991, \mn@doi [\apj] {10.1086/170078}, \href
  {https://ui.adsabs.harvard.edu/abs/1991ApJ...373..592L} {373, 592}

\bibitem[\protect\citeauthoryear{{Link}, {Epstein}  \& {Lattimer}}{{Link}
  et~al.}{1999}]{link:1999iek}
{Link} B.,  {Epstein} R.~I.,   {Lattimer} J.~M.,  1999, \mn@doi [\prl]
  {10.1103/PhysRevLett.83.3362}, \href
  {https://ui.adsabs.harvard.edu/abs/1999PhRvL..83.3362L} {83, 3362}

\bibitem[\protect\citeauthoryear{Ljung}{Ljung}{1986}]{Ljung:1986:SIT:21413}
Ljung L.,  1986, System Identification: Theory for the User.
Prentice-Hall, Inc., Upper Saddle River, NJ, USA

\bibitem[\protect\citeauthoryear{Ljung}{Ljung}{2013}]{Ljung2013}
Ljung L.,  2013, \mn@doi [Journal of Control, Automation and Electrical
  Systems] {10.1007/s40313-013-0004-7}, 24, 3

\bibitem[\protect\citeauthoryear{{Lower} et~al.,}{{Lower}
  et~al.}{2020}]{2020MNRAS.tmp..578L}
{Lower} M.~E.,  et~al., 2020, \mn@doi [\mnras] {10.1093/mnras/staa615}, \href
  {https://ui.adsabs.harvard.edu/abs/2020MNRAS.tmp..578L} {}

\bibitem[\protect\citeauthoryear{Lyne \& Graham-Smith}{Lyne \&
  Graham-Smith}{2012}]{lyne_graham-smith_2012}
Lyne A.,  Graham-Smith F.,  2012, Pulsar Astronomy, 4 edn.
Cambridge Astrophysics, Cambridge University Press,
  \mn@doi{10.1017/CBO9780511844584}

\bibitem[\protect\citeauthoryear{{Lyne}, {Hobbs}, {Kramer}, {Stairs}  \&
  {Stappers}}{{Lyne} et~al.}{2010}]{lyne:2010pds}
{Lyne} A.,  {Hobbs} G.,  {Kramer} M.,  {Stairs} I.,   {Stappers} B.,  2010,
  \mn@doi [Science] {10.1126/science.1186683}, \href
  {https://ui.adsabs.harvard.edu/abs/2010Sci...329..408L} {329, 408}

\bibitem[\protect\citeauthoryear{MathWorks}{MathWorks}{2019}]{MatlabEconometricsToolbox}
MathWorks 2019, MATLAB Econometrics Toolbox

\bibitem[\protect\citeauthoryear{{Matsakis}, {Taylor}  \& {Eubanks}}{{Matsakis}
  et~al.}{1997}]{matsakis:1997lsd}
{Matsakis} D.~N.,  {Taylor} J.~H.,   {Eubanks} T.~M.,  1997, \aap, \href
  {https://ui.adsabs.harvard.edu/abs/1997A&A...326..924M} {326, 924}

\bibitem[\protect\citeauthoryear{{Melatos}}{{Melatos}}{2012}]{melatos:2012ped}
{Melatos} A.,  2012, \mn@doi [\apj] {10.1088/0004-637X/761/1/32}, \href
  {https://ui.adsabs.harvard.edu/abs/2012ApJ...761...32M} {761, 32}

\bibitem[\protect\citeauthoryear{Melatos \& Link}{Melatos \&
  Link}{2014}]{Melatos2014}
Melatos A.,  Link B.,  2014, \mn@doi [\mnras] {10.1093/mnras/stt1828}, 437, 21

\bibitem[\protect\citeauthoryear{{Melatos} \& {Mastrano}}{{Melatos} \&
  {Mastrano}}{2016}]{melatos:2016aaf}
{Melatos} A.,  {Mastrano} A.,  2016, \mn@doi [\apj]
  {10.3847/0004-637X/818/1/49}, \href
  {https://ui.adsabs.harvard.edu/abs/2016ApJ...818...49M} {818, 49}

\bibitem[\protect\citeauthoryear{{Melatos} \& {Payne}}{{Melatos} \&
  {Payne}}{2005}]{melatos:2005yup}
{Melatos} A.,  {Payne} D.~J.~B.,  2005, \mn@doi [\apj] {10.1086/428600}, \href
  {https://ui.adsabs.harvard.edu/abs/2005ApJ...623.1044M} {623, 1044}

\bibitem[\protect\citeauthoryear{Melatos \& Peralta}{Melatos \&
  Peralta}{2007}]{Melatos_2007}
Melatos A.,  Peralta C.,  2007, \mn@doi [The Astrophysical Journal]
  {https://doi.org/10.1086/518598}, 662, L99

\bibitem[\protect\citeauthoryear{{Melatos} \& {Peralta}}{{Melatos} \&
  {Peralta}}{2010}]{melatos:2010bvy}
{Melatos} A.,  {Peralta} C.,  2010, \mn@doi [\apj]
  {10.1088/0004-637X/709/1/77}, \href
  {https://ui.adsabs.harvard.edu/abs/2010ApJ...709...77M} {709, 77}

\bibitem[\protect\citeauthoryear{{Melatos}, {Howitt}  \& {Fulgenzi}}{{Melatos}
  et~al.}{2018}]{Melatos:2018lsd}
{Melatos} A.,  {Howitt} G.,   {Fulgenzi} W.,  2018, \mn@doi [\apj]
  {10.3847/1538-4357/aad228}, \href
  {https://ui.adsabs.harvard.edu/abs/2018ApJ...863..196M} {863, 196}

\bibitem[\protect\citeauthoryear{Mendell}{Mendell}{1991a}]{Mendell1991a}
Mendell G.,  1991a, \mn@doi [\apj] {10.1086/170609}, 380, 515

\bibitem[\protect\citeauthoryear{Mendell}{Mendell}{1991b}]{Mendell1991}
Mendell G.,  1991b, \mn@doi [\apj] {10.1086/170610}, 380, 530

\bibitem[\protect\citeauthoryear{Miller et~al.,}{Miller
  et~al.}{2019}]{Miller_2019}
Miller M.~C.,  et~al., 2019, \mn@doi [The Astrophysical Journal]
  {10.3847/2041-8213/ab50c5}, 887, L24

\bibitem[\protect\citeauthoryear{Mukherjee, Messenger  \& Riles}{Mukherjee
  et~al.}{2018}]{Mukherjee:2018rim}
Mukherjee A.,  Messenger C.,   Riles K.,  2018, \mn@doi [Phys. Rev. D]
  {10.1103/PhysRevD.97.043016}, 97, 043016

\bibitem[\protect\citeauthoryear{{Namkham}, {Jaroenjittichai}  \&
  {Johnston}}{{Namkham} et~al.}{2019}]{2019MNRAS.487.5854N}
{Namkham} N.,  {Jaroenjittichai} P.,   {Johnston} S.,  2019, \mn@doi [\mnras]
  {10.1093/mnras/stz1671}, \href
  {https://ui.adsabs.harvard.edu/abs/2019MNRAS.487.5854N} {487, 5854}

\bibitem[\protect\citeauthoryear{{Parthasarathy} et~al.,}{{Parthasarathy}
  et~al.}{2019}]{2019MNRAS.489.3810P}
{Parthasarathy} A.,  et~al., 2019, \mn@doi [\mnras] {10.1093/mnras/stz2383},
  \href {https://ui.adsabs.harvard.edu/abs/2019MNRAS.489.3810P} {489, 3810}

\bibitem[\protect\citeauthoryear{{Parthasarathy} et~al.,}{{Parthasarathy}
  et~al.}{2020}]{parthasarathy:2020wel}
{Parthasarathy} A.,  et~al., 2020, arXiv e-prints, \href
  {https://ui.adsabs.harvard.edu/abs/2020arXiv200313303P} {p. arXiv:2003.13303}

\bibitem[\protect\citeauthoryear{Peralta, Melatos, Giacobello  \& Ooi}{Peralta
  et~al.}{2005}]{Peralta2005}
Peralta C.,  Melatos A.,  Giacobello M.,   Ooi A.,  2005, \mn@doi [\apj]
  {10.1086/497899}, 635, 1224

\bibitem[\protect\citeauthoryear{{Price}, {Link}, {Shore}  \& {Nice}}{{Price}
  et~al.}{2012}]{2012MNRAS.426.2507P}
{Price} S.,  {Link} B.,  {Shore} S.~N.,   {Nice} D.~J.,  2012, \mn@doi [\mnras]
  {10.1111/j.1365-2966.2012.21863.x}, \href
  {https://ui.adsabs.harvard.edu/abs/2012MNRAS.426.2507P} {426, 2507}

\bibitem[\protect\citeauthoryear{{Priymak}, {Melatos}  \& {Payne}}{{Priymak}
  et~al.}{2011}]{priymak:2011try}
{Priymak} M.,  {Melatos} A.,   {Payne} D.~J.~B.,  2011, \mn@doi [\mnras]
  {10.1111/j.1365-2966.2011.19431.x}, \href
  {https://ui.adsabs.harvard.edu/abs/2011MNRAS.417.2696P} {417, 2696}

\bibitem[\protect\citeauthoryear{Raaijmakers et~al.,}{Raaijmakers
  et~al.}{2019}]{Raaijmakers_2019}
Raaijmakers G.,  et~al., 2019, \mn@doi [The Astrophysical Journal]
  {10.3847/2041-8213/ab451a}, 887, L22

\bibitem[\protect\citeauthoryear{Rauch, Tung  \& Striebel}{Rauch
  et~al.}{1965}]{rts_smoother_paper}
Rauch H.~E.,  Tung F.,   Striebel C.~T.,  1965, \mn@doi [AIAA Journal]
  {10.2514/3.3166}, 3, 1445

\bibitem[\protect\citeauthoryear{{Reisenegger}}{{Reisenegger}}{1993}]{Reisenegger:1993aab}
{Reisenegger} A.,  1993, \mn@doi [Journal of Low Temperature Physics]
  {10.1007/BF00681873}, \href
  {https://ui.adsabs.harvard.edu/abs/1993JLTP...92...77R} {92, 77}

\bibitem[\protect\citeauthoryear{{Riles}}{{Riles}}{2013}]{2013PrPNP..68....1R}
{Riles} K.,  2013, \mn@doi [Progress in Particle and Nuclear Physics]
  {10.1016/j.ppnp.2012.08.001}, \href
  {https://ui.adsabs.harvard.edu/abs/2013PrPNP..68....1R} {68, 1}

\bibitem[\protect\citeauthoryear{Riley et~al.,}{Riley
  et~al.}{2019}]{Riley_2019}
Riley T.~E.,  et~al., 2019, \mn@doi [The Astrophysical Journal]
  {10.3847/2041-8213/ab481c}, 887, L21

\bibitem[\protect\citeauthoryear{{Romanova}, {Ustyugova}, {Koldoba}  \&
  {Lovelace}}{{Romanova} et~al.}{2004}]{romanova2004khj}
{Romanova} M.~M.,  {Ustyugova} G.~V.,  {Koldoba} A.~V.,   {Lovelace} R.~V.~E.,
  2004, \mn@doi [\apjl] {10.1086/426586}, \href
  {https://ui.adsabs.harvard.edu/abs/2004ApJ...616L.151R} {616, L151}

\bibitem[\protect\citeauthoryear{{Romanova}, {Kulkarni}  \&
  {Lovelace}}{{Romanova} et~al.}{2008}]{romanova:2008ieu}
{Romanova} M.~M.,  {Kulkarni} A.~K.,   {Lovelace} R. V.~E.,  2008, \mn@doi
  [\apjl] {10.1086/527298}, \href
  {https://ui.adsabs.harvard.edu/abs/2008ApJ...673L.171R} {673, L171}

\bibitem[\protect\citeauthoryear{R\"oßler}{R\"oßler}{2010}]{10.2307/41062628}
R\"oßler A.,  2010, SIAM Journal on Numerical Analysis, 48, 922

\bibitem[\protect\citeauthoryear{{Ruderman}, {Zhu}  \& {Chen}}{{Ruderman}
  et~al.}{1998}]{ruderman:1998ids}
{Ruderman} M.,  {Zhu} T.,   {Chen} K.,  1998, \mn@doi [\apj] {10.1086/305026},
  \href {https://ui.adsabs.harvard.edu/abs/1998ApJ...492..267R} {492, 267}

\bibitem[\protect\citeauthoryear{{Sedrakian} \& {Hairapetian}}{{Sedrakian} \&
  {Hairapetian}}{2002}]{sedrakian:2002vbf}
{Sedrakian} D.~M.,  {Hairapetian} M.~V.,  2002, \mn@doi [Astrophysics]
  {10.1023/A:1021859113721}, \href
  {https://ui.adsabs.harvard.edu/abs/2002Ap.....45..470S} {45, 470}

\bibitem[\protect\citeauthoryear{{Seveso}, {Pizzochero}, {Grill}  \&
  {Haskell}}{{Seveso} et~al.}{2016}]{Seveso:2016wqd}
{Seveso} S.,  {Pizzochero} P.~M.,  {Grill} F.,   {Haskell} B.,  2016, \mn@doi
  [\mnras] {10.1093/mnras/stv2579}, \href
  {https://ui.adsabs.harvard.edu/abs/2016MNRAS.455.3952S} {455, 3952}

\bibitem[\protect\citeauthoryear{Shannon \& Cordes}{Shannon \&
  Cordes}{2010}]{Shannon2010}
Shannon R.~M.,  Cordes J.~M.,  2010, \mn@doi [\apj]
  {10.1088/0004-637X/725/2/1607}, 725, 1607

\bibitem[\protect\citeauthoryear{Shumway \& Stoffer}{Shumway \&
  Stoffer}{1982}]{Shumway1982}
Shumway R.~H.,  Stoffer D.~S.,  1982, \mn@doi [Journal of Time Series Analysis]
  {10.1111/j.1467-9892.1982.tb00349.x}, 3, 253

\bibitem[\protect\citeauthoryear{{Sidery} \& {Alpar}}{{Sidery} \&
  {Alpar}}{2009}]{sidery:2009kds}
{Sidery} T.,  {Alpar} M.~A.,  2009, \mn@doi [\mnras]
  {10.1111/j.1365-2966.2009.15575.x}, \href
  {https://ui.adsabs.harvard.edu/abs/2009MNRAS.400.1859S} {400, 1859}

\bibitem[\protect\citeauthoryear{{Sidery}, {Passamonti}  \&
  {Andersson}}{{Sidery} et~al.}{2010}]{2010MNRAS.405.1061S}
{Sidery} T.,  {Passamonti} A.,   {Andersson} N.,  2010, \mn@doi [\mnras]
  {10.1111/j.1365-2966.2010.16497.x}, \href
  {https://ui.adsabs.harvard.edu/abs/2010MNRAS.405.1061S} {405, 1061}

\bibitem[\protect\citeauthoryear{{Stairs} et~al.,}{{Stairs}
  et~al.}{2019}]{stairs:2019jfs}
{Stairs} I.~H.,  et~al., 2019, \mn@doi [\mnras] {10.1093/mnras/stz647}, \href
  {https://ui.adsabs.harvard.edu/abs/2019MNRAS.485.3230S} {485, 3230}

\bibitem[\protect\citeauthoryear{{Sun}, {Melatos}, {Suvorova}, {Moran}  \&
  {Evans}}{{Sun} et~al.}{2018}]{2018PhRvD..97d3013S}
{Sun} L.,  {Melatos} A.,  {Suvorova} S.,  {Moran} W.,   {Evans} R.~J.,  2018,
  \mn@doi [\prd] {10.1103/PhysRevD.97.043013}, \href
  {https://ui.adsabs.harvard.edu/abs/2018PhRvD..97d3013S} {97, 043013}

\bibitem[\protect\citeauthoryear{{Suvorova}, {Sun}, {Melatos}, {Moran}  \&
  {Evans}}{{Suvorova} et~al.}{2016}]{2016PhRvD..93l3009S}
{Suvorova} S.,  {Sun} L.,  {Melatos} A.,  {Moran} W.,   {Evans} R.~J.,  2016,
  \mn@doi [\prd] {10.1103/PhysRevD.93.123009}, \href
  {https://ui.adsabs.harvard.edu/abs/2016PhRvD..93l3009S} {93, 123009}

\bibitem[\protect\citeauthoryear{{Suvorova}, {Clearwater}, {Melatos}, {Sun},
  {Moran}  \& {Evans}}{{Suvorova} et~al.}{2017}]{2017PhRvD..96j2006S}
{Suvorova} S.,  {Clearwater} P.,  {Melatos} A.,  {Sun} L.,  {Moran} W.,
  {Evans} R.~J.,  2017, \mn@doi [\prd] {10.1103/PhysRevD.96.102006}, \href
  {https://ui.adsabs.harvard.edu/abs/2017PhRvD..96j2006S} {96, 102006}

\bibitem[\protect\citeauthoryear{{Ushomirsky}, {Cutler}  \&
  {Bildsten}}{{Ushomirsky} et~al.}{2000}]{ushomirsky:2000bvc}
{Ushomirsky} G.,  {Cutler} C.,   {Bildsten} L.,  2000, \mn@doi [\mnras]
  {10.1046/j.1365-8711.2000.03938.x}, \href
  {https://ui.adsabs.harvard.edu/abs/2000MNRAS.319..902U} {319, 902}

\bibitem[\protect\citeauthoryear{{Warszawski} \& {Melatos}}{{Warszawski} \&
  {Melatos}}{2011}]{Warszawski:2011ldo}
{Warszawski} L.,  {Melatos} A.,  2011, \mn@doi [\mnras]
  {10.1111/j.1365-2966.2011.18803.x}, \href
  {https://ui.adsabs.harvard.edu/abs/2011MNRAS.415.1611W} {415, 1611}

\bibitem[\protect\citeauthoryear{{Warszawski} \& {Melatos}}{{Warszawski} \&
  {Melatos}}{2013}]{melatos:2013pwe}
{Warszawski} L.,  {Melatos} A.,  2013, \mn@doi [\mnras] {10.1093/mnras/sts108},
  \href {https://ui.adsabs.harvard.edu/abs/2013MNRAS.428.1911W} {428, 1911}

\bibitem[\protect\citeauthoryear{{Yakovlev}, {Levenfish}  \&
  {Shibanov}}{{Yakovlev} et~al.}{1999}]{1999PhyU...42..737Y}
{Yakovlev} D.~G.,  {Levenfish} K.~P.,   {Shibanov} Y.~A.,  1999, \mn@doi
  [Physics Uspekhi] {10.1070/PU1999v042n08ABEH000556}, \href
  {https://ui.adsabs.harvard.edu/abs/1999PhyU...42..737Y} {42, 737}

\bibitem[\protect\citeauthoryear{de Jong \& Shephard}{de~Jong \&
  Shephard}{1995}]{dejong:1995yex}
de Jong P.,  Shephard N.,  1995, Biometrika, 82, 339

\bibitem[\protect\citeauthoryear{{de Kool} \& {Anzer}}{{de Kool} \&
  {Anzer}}{1993}]{deKool:1993wvd}
{de Kool} M.,  {Anzer} U.,  1993, \mn@doi [\mnras] {10.1093/mnras/262.3.726},
  \href {https://ui.adsabs.harvard.edu/abs/1993MNRAS.262..726D} {262, 726}

\bibitem[\protect\citeauthoryear{{van Eysden} \& {Melatos}}{{van Eysden} \&
  {Melatos}}{2010}]{2010MNRAS.409.1253V}
{van Eysden} C.~A.,  {Melatos} A.,  2010, \mn@doi [\mnras]
  {10.1111/j.1365-2966.2010.17387.x}, \href
  {https://ui.adsabs.harvard.edu/abs/2010MNRAS.409.1253V} {409, 1253}

\bibitem[\protect\citeauthoryear{van Eysden \& Melatos}{van Eysden \&
  Melatos}{2013}]{van_Eysden_2013}
van Eysden C.~A.,  Melatos A.,  2013, \mn@doi [Journal of Fluid Mechanics]
  {https://ui.adsabs.harvard.edu/link_gateway/2013JFM...729..180V/doi:10.1017/jfm.2013.294},
  729, 180

\makeatother
\end{thebibliography}

\appendix
\section{Analytic Solution of the Langevin equations (\ref{EQ:CRUST_EQ_OF_MOTION}) and (\ref{EQ:SUPERFLUID_EQ_OF_MOTION})}
\label{app:sec:analytic_solution}
The pair of coupled linear inhomogeneous equations in~(\ref{EQ:CRUST_EQ_OF_MOTION}) and (\ref{EQ:SUPERFLUID_EQ_OF_MOTION}) have an analytic 
solution. We break our solution into two parts $\omgc(t) = \omgc^P(t) + \omgc^H(t)$, 
where superscripts $H$ and $P$ indicate homogeneous and particular solutions respectively. 
An analogous expression holds for $\omgs(t)$.

The homogeneous solutions are found by solving (\ref{EQ:CRUST_EQ_OF_MOTION}) and (\ref{EQ:SUPERFLUID_EQ_OF_MOTION}) with $N_{\rm c,s}=0$ and $\xi_{\rm c,s}=0$. The result is
% \begin{widetext}
\begin{align}
\omgc^H(t) &= \frac{1}{\taus + \tauc}\left[\omgc(0)(\tauc+\taus e^{-t/\tau}) + \omgs(0)(\taus - \taus e^{- t/\tau})\right]\label{eq:analytic:omgc_h}\\
\omgs^H(t) &= \frac{1}{\taus + \tauc}\left[\omgc(0)(\tauc-\tauc e^{-t/\tau}) + \omgs(0)(\taus + \tauc e^{-t/\tau})\right]\label{eq:analytic:omgs_h}\\
\tau&=\frac{\tauc\taus}{\tauc+\taus},
\end{align}
where $\tau$ is the relaxation time of the system.
The particular solution includes an integration over the stochastic history of the random variables $\xi_{\rm c,s}(t)$
\begin{align}
\nonumber\Omega_c^P(t) = &\frac{1}{\tauc + \taus} \left\{\frac{\Nc}{\Ic}\tauc t + \frac{\Ns}{\Is} \taus t + \right.\\
\nonumber&\left.\frac{\Nc}{\Ic}\taus\tau\left(1 - e^{-t/\tau}\right) -\frac{\Ns}{\Is} \taus\tau\left(1 - e^{- t/\tau}\right)+\right.\\
\nonumber&\left. \int_0^{t} dt'\;\left[\frac{\xi_{\rm c}(t')}{\Ic}\left(\tauc + \taus \expdiff\right) +\right.\right.\\
&\left.\left. \frac{\xi_{\rm s}(t')}{I_s}\left(\taus -\taus\expdiff\right)\right]\right\}\label{eq:analytic:omgc_p}\\
\nonumber\Omega_s^P(t)=&\frac{1}{\tauc+\taus}\left\{\frac{\Nc}{\Ic}\tauc t + \frac{\Ns}{\Is} \taus t -\right.\\
\nonumber&\left. \tauc \frac{\Nc}{\Ic}\tau\left(1 - e^{-t/\tau}\right) + \frac{\Ns}{\Is} \tauc\tau\left(1 - e^{-t/\tau}\right)+\right.\\
\nonumber&\left.\int_0^{t}dt'\;\left[\frac{\xi_{\rm c}(t')}{\Ic}\left(\tauc - \tauc\expdiff\right) + \right.\right.\\
&\left.\left.\frac{\xi_{\rm s}(t')}{\Is}\left(\taus + \tauc \expdiff\right)\right]\right\}\label{eq:analytic:omgs_p}.
\end{align}
It is worthwhile to note that in the limit with $t\gg\tauc,\;\taus$, the difference between the two spins converges to
\begin{align}
\label{eq:analytic_solution:ang_vel_diff}
\omgc(t)-\omgs(t) &= \tau\left(\frac{\Nc}{\Ic} - \frac{\Ns}{\Is}\right) + \textrm{stochastic terms}
\end{align}
and the ensemble-averaged spin down is the same for both components:
\begin{align}
\langle \dot\Omega_{\rm c}\rangle &= \langle\dot\Omega_{\rm s}\rangle\\
&= \frac{1}{\tauc + \taus}\left(\tauc\frac{\Nc}{\Ic} + \taus\frac{\Ns}{\Is} \right).
\end{align}

\section{Expectation-maximization algorithm: overview}
\label{sec:em_overview_for_our_case}
In this appendix we discuss the application of the expectation-maximization algorithm to our specific problem. We first present the full form of $p(\bm\theta,\bm x| \bm y)$ in Appendix~\ref{sssec:infer:em:likelihood}. We then discuss the E-step in Appendix~\ref{sssec:em:estep} and M-step in Appendix~\ref{sssec:em:mstep}, along with the practical implementation of the algorithm used in the rest of this paper in Appendix~\ref{sssec:em:prac_imp}. In Appendix~\ref{app:em_implementation}, we show in detail the explicit formulae used in carrying out the E-step and the M-step.
\subsection{Likelihood}
\label{sssec:infer:em:likelihood}

We use a Gaussian of the form 
\begin{align}
\label{eq:inference:log_likelihood}
\nonumber    \log{p(\bs x, \bs\theta | \bs y)} = &- \frac{1}{2}(\bs x_0 - \bs \mu)^T \Sigma_{00}^{-1}(\bs x_0 - \bs \mu)\\
\nonumber    & - \frac{1}{2}\sum_{k=1}^{N_t}(\bs x_{k} - \transition\bs x_{k-1}-\torques)^T\processcovar^{-1}(\bs x_{k} - \transition\bs x_{k-1}-\torques{})\\
\nonumber    &  - \frac{1}{2}\sum_{k=1}^{N_t}(\bs y_{k} - \emission\bs x_{k})^T\measurecovar^{-1}(\bs y_{k} - \emission\bs x_{k})\\
    &-\frac{1}{2}\log |\Sigma_{00}| -\frac{N_t}{2}\log |\measurecovar|-\frac{N_t}{2}\log |\processcovar| + \textrm{const},
\end{align}
where $N_t$ is the number of measurements, and $|\bs R|$ and $|\bs Q|$ are unsigned determinants.
The first line in (\ref{eq:inference:log_likelihood}) gives the probability of the initial state $\bs x_0$, where $\Sigma_{00}=\langle \bs x_0 \bs x_0^T\rangle$ is the covariance matrix of the parameters in the initial state. The second line describes the probability that the Wiener process (here, spin wandering) makes a jump of size $\Delta\bs{W}_k = \bs x_{k+1} - \transition\bs{x}_k - \torques$. The third line in (\ref{eq:inference:log_likelihood}) describes the probability that the measurement noise takes the value $\bs{u}_k = \bs{y}_k - \bs B \bs{x}_k$. The final line is included for normalization, with ``const'' being a constant normalization factor that is independent of the parameters.

We can now calculate $E_p[\log p(\bs\theta, \bs x | \bs y)]$, which is defined in (\ref{eq:expectation_of_posterior}). As discussed above, we treat $\bs x$ as the random variable here. we can write $E_p[\log p(\bs\theta, \bs x | \bs y)]$ in terms of state estimates and state covariances, which we denote as
\begin{align}
\bs{\hat{x}}_{k|N_t}&=E_p(\bs x_k | \bs y, \bs\theta_p)\label{eq:expectation_of_x}\\
\bs\Sigma_{k, k'|N_t}&=E_p[(\bs x_k - \bs{\hat{x}}_{k|N_t})(\bs x_{k'} - \bs{\hat{x}}_{k'|N_t})^T| \bs y, \bs\theta_p].\label{eq:covariance_of_x}
\end{align}
The notation $\bs{\hat{x}}_{k|N_t}$ indicates an estimate of the states at time-step $k$ using all data through time-step $N_t$. The explicit form of $E_p[\log p(\bs\theta, \bs x | \bs y)]$, in terms of $\bs{\hat{x}}_{k|N_t}$ and $\bs\Sigma_{k, k'|N_t}$ is shown in (\ref{eq:expected_log_likelihood_old}).

Equations (\ref{eq:expectation_of_x}) and (\ref{eq:covariance_of_x}) can be evaluated using a forward-backward RTS smoother~\citep{rts_smoother_paper}, using $\bs\theta_p$ for the elements of $\transition$, $\torques$, $\processcovar$. The forward steps of the smoother are the same as the Kalman filter and are shown in (\ref{eq:estep:forward1})--(\ref{eq:estep:sig_k_k-1}). The backwards steps are shown in (\ref{eq:estep:backward1})--(\ref{eq:estep:backwardN}).

We can also maximize $E_p[\log p(\bs\theta, \bs x | \bs y)]$, in (\ref{eq:expected_log_likelihood_old}), over the parameters $\bs\theta$ by setting 

\begin{align}
\frac{\partial E_p[\log p(\bs\theta, \bs x | \bs y)]}{\partial\bs\theta} = 0.
\end{align}
This yields a set of maximum-likelihood expressions for a new value of $\bs\theta$ in terms of $\bs{\hat{x}}_{k|N_t}$ and $\bs\Sigma_{k, k'|N_t}$ (estimated using an ``old'' version of the parameters, $\bs\theta_p$). These maximum-likelihood expressions are shown in (\ref{eq:mlef})--(\ref{eq:mleq}). We are now ready to discuss how to implement the expectation-maximization algorithm. 

\subsection{E-step}
\label{sssec:em:estep}
In the E-step we run the RTS-smoother, using an ``old'' set of parameter estimates $\bs\theta_p$ (where $p$ tracks which iteration of the algorithm we are on) to estimate $\bs{\hat{x}}_{k|N_t}$ and $\bs\Sigma_{k, k'|N_t}$.

\subsection{M-step}
\label{sssec:em:mstep}
In the M-step we use $\bs{\hat{x}}_{k|N_t}$ and $\bs\Sigma_{k, k'|N_t}$ from the E-step to update our estimates of $\bs\theta$ using the maximum likelihood expressions in (\ref{eq:mlef})--(\ref{eq:mleq}). This produces a new set of parameters $\bs\theta_{p+1}$ that are passed to the E-step of the next iteration.

\subsection{Practical implementation}
\label{sssec:em:prac_imp}
To summarize, we estimate the states and their covariances, given an estimate of the parameters in the E-step. We then maximize the likelihood over the parameters in terms of those state estimates.

The expectation-maximization algorithm converges to a stationary point of the marginal posterior distribution $p(\bs\theta | \bs y)$, assuming one exists. If multiple stationary points exist then the algorithm converges to a local maximum. Therefore, we run the algorithm many times with different, random starting guesses for $\bs\theta$ and then take the maximum likelihood recovery across all of those runs. This is feasible because each iteration is computationally cheap and the algorithm typically converges within $\mathcal O(10)$ iterations.

In practice, we use the \texttt{ssm} package in the MATLAB econometrics toolbox to perform the parameter estimation, which is a maximum-likelihood method similar to the expectation-maximization algorithm presented here~\citep{MatlabEconometricsToolbox}.

\section{Expectation-maximization algorithm: key formulae}
\label{app:em_implementation}
In this appendix we present the implementation of the expectation-maximization algorithm at a more practical level. We explicitly calculate $E_p[\log p(\bs x, \bs \theta | \bs y)]$ in terms of the state estimates and covariances, we show how to find the parameter estimates which maximize $E_p[\log p(\bs x, \bs \theta | \bs y)]$, and finally we show how to estimate the states and covariances using an RTS smoother.

We calculate $E_p[\log p(\bs x, \bs \theta | \bs y)]$ defined in~(\ref{eq:expectation_of_posterior}), treating $\bs x$ as a random variable, in terms of the state estimates and covariances defined in~(\ref{eq:expectation_of_x}) and (\ref{eq:covariance_of_x}). Taking the expectation of~(\ref{eq:inference:log_likelihood}), we find
\begin{align}
\nonumber E_p\left[\log{p(\bs x, \bs\theta | \bs y)}\right] = &- \frac{1}{2}\textrm{tr}\left\{\Sigma_{00}^{-1}\left[\bs{\Sigma}_{0,0|N_t} + (\hat{\bs{x}}_0 - \bs\mu) (\hat{\bs{x}}_0 - \bs\mu)^T\right]\right\}\\
\nonumber& - \frac{1}{2}\textrm{tr}\left\{\processcovar^{-1}\left[\bs \Gamma_2 - \bs \Gamma_4\transition^T - \transition \bs \Gamma_4^{T} + \transition\bs \Gamma_3\transition^T\right.\right.\\
\nonumber&\left.\left. - \bs\gamma_2\torques^T + \transition\bs\gamma_1\torques^T - \torques\bs\gamma_2^T + \torques\bs\gamma_1^T\transition^T \right.\right.\\\nonumber&\left.\left.+ N_t\torques \torques^T\right]\right\}\\
\nonumber& - \frac{1}{2}\textrm{tr}\left\{\measurecovar^{-1}\sum_{k=1}^{N_t}\left[(\bs y_k - \emission\hat{\bs{x}}_k)(\bs y_k - \emission\hat{\bs{x}}_k)^T\right]\right\}\\
&-\frac{1}{2}\log |\Sigma_{00}|  -\frac{N_t}{2}\log|\processcovar|-\frac{N_t}{2}\log|\measurecovar| + \textrm{const}
\label{eq:expected_log_likelihood_old}
\end{align}
where we define the following useful statistics:
\begin{align}
\label{eq:gammas_exp1}{\bs{\gamma}}_1 &= \sum_{k=1}^{N_t} \hat{\bs{x}}_{k-1|N_t}\\
{\bs{\gamma}}_2 &= \sum_{k=1}^{N_t} \hat{\bs{x}}_{k|N_t}\\
{\bs{\Gamma}}_2 &= \sum_{k=1}^{N_t} \left(\bs{\Sigma}_{k,k | N_t} + \hat{\bs{x}}_{k|N_t}\hat{\bs{x}}_{k|N_t}^T\right)\\
{\bs{\Gamma}}_3 &= \sum_{k=1}^{N_t} \left(\bs{\Sigma}_{k-1, k-1| N_t} + \hat{\bs{x}}_{k-1|N_t}\hat{\bs{x}}_{k-1|N_t}^T\right)\\
\label{eq:gammas_expN}{\bs{\Gamma}}_4 &= \sum_{k=1}^{N_t} \left(\bs{\Sigma}_{k, k-1 | N_t} +  \hat{\bs{x}}_{k|N_t}\hat{\bs{x}}_{k-1|N_t}^T\right).
\end{align}

We then find the parameter estimates that maximize  $E_p[\log p(\bs x, \bs \theta | \bs y)]$. For example, we solve
\begin{align}
\frac{\partial E_p[\log p(\bs x, \bs \theta | \bs y)]}{\partial\transition} = 0,
\end{align}
for the elements of $\transition$ in terms of the statistics in~(\ref{eq:gammas_exp1})--(\ref{eq:gammas_expN}). Solving for $\transition,\,\processcovar,$ and $\torques$, yields
\begin{align} 
\hat{\bs{F}} =& \left(\bs{\Gamma}_4 - \frac{1}{N_t}\bs\gamma_2\bs{\gamma}_1^T\right)\left(\bs{\Gamma}_3 - \frac{1}{N_t}\bs\gamma_1\bs\gamma_1^T\right)^{-1}\label{eq:mlef}\\
\hat{\torques} =&\, \frac{1}{N_t}\left(\bs{\gamma}_2 - \hat{\transition} \bs{\gamma}_1\right) \label{eq:mlea}\\
\hat{\bs{Q}} =&\, \frac{1}{N_t}\left(\bs{\Gamma}_2 - \hat{\transition}\bs{\Gamma}_4^T - \bs{\Gamma}_4\hat{\transition}^T \nonumber\right.\\&-\bs{\gamma}_2 \hat{\torques}^T - \hat{\torques}\bs{\gamma}_2^T + \hat{\transition}\bs{\gamma}_1\hat{\torques}^T \nonumber\\&\left.+\hat{\torques}\bs{\gamma}_1^T\hat{\transition}^T + \hat{\transition}\bs{\Gamma}_3\hat{\transition}^T + N_t\hat{\torques} \hat{\torques}^T\right).\label{eq:mleq}
\end{align}
There is some intuition to be gained by considering these first two expressions individually. In the first expression, we can rearrange by right-hand multiplying by $\bs\Gamma_3 - \gamma_1\gamma_1^T / N_t$. We can then read this expression as an indication that the transition matrix applied to the variance of each state should give the covariance between that state and its subsequent state. Intuitively, this makes sense, and this fact is also used below in (\ref{eq:estep:sig_k_k-1}). In the second expression, the average difference between the current state and the previous state multiplied by the transition matrix gives the linear, constant torque. This is essentially a restatement of (\ref{eq:inference:state_transition_equation}), averaged over all available data.

To calculate the updated statistics in (\ref{eq:mlef})--(\ref{eq:mleq}), we need to estimate $\bs{\hat{x}}_{k|N_t}$ and $\bs\Sigma_{k, k'|N_t}$ (and hence the statistics in [\ref{eq:gammas_exp1}]--[\ref{eq:gammas_expN}]) from $\bs{y}$. We do this using an RTS smoother, which is a forward-backward algorithm that uses all $N_t$ observations to estimate each $\bs x_k$. The forward steps calculate $\hat{\bs{x}}_{k|k}$, $\bs\Sigma_{k,k|k}$, and $\bs \Sigma_{k,k-1|k}$ and are given in~\citep{Shumway1982}. We first project the states, $\bs x$ forward one step,
\begin{align}
\label{eq:estep:forward1}
\hat{\bs{x}}_{k+1 | k} &= \hat{\bs{F}} \hat{\bs{x}}_{k | k} + \hat{\torques}\\
\bs{\Sigma}_{k+1,k+1 | k} &= \hat{\bs{F}} \bs{\Sigma}_{k,k | k} \hat{\bs{F}}^T + \hat{\bs{Q}}.
\end{align}
We then correct those states using our measurements, $\bs y$,
\begin{align}
\bs{K}_k &= \bs{\Sigma}_{k,k | k-1}\bs{B}^T\left(\bs{B} \bs{\Sigma}_{k,k | k-1} \bs{B}^T + \bs{R}\right)^{-1}\\
\bs{\hat{x}}_{k|k} &= \hat \bx_{k | k-1} + \bs{K}_k (\hat \by_k - \bs B\hat \bx_{k | k-1})\\
\bs{\Sigma}_{k,k | k} &= \bs{\Sigma}_{k,k | k-1} - \bs{K}_k \bs{\Sigma}_{k,k | k-1} \bs{K}_k^T\\
\bs{\Sigma}_{k, k-1 | k} &= \left(\bs{I} - \bs{K}_k \bs{B}\right)\hat{\bs{F}}\bs{\Sigma}_{k-1,k-1 | k-1}. \label{eq:estep:sig_k_k-1}
\end{align}
We can use a set of backwards recursions to estimate the state at each time step using all of the data. The backwards steps then use $\hat{\bs{x}}_{N_t|N_t}$ to update $\hat{\bs{x}}_{N_{t}-1|N_{t}-1}$ and so on. The back-tracking steps are given by~\citep{Shumway1982}
\begin{align}
\label{eq:estep:backward1}\hat{\bs{x}}_{k-1 | N_t} &= \hat{\bx}_{k-1 | k-1} + \bs{M}_k\left[\hat{\bx}_{k | N_t} - \hat{\bx}_{k | k-1} \right]\\
\bs{\Sigma}_{k-1,k-1 | N_t} &= \hat{\bs{\Sigma}}_{k-1,k-1 | k-1} + \bs{M}_k\left[\bs{\Sigma}_{k,k | N_t} - \bs{\Sigma}_{k,k | k-1} \right]\bs{M}_k^T\\
\bs{M}_k &= \bs{\Sigma}_{k-1, k-1 | k-1} \hat{\bs{F}}^T\bs{\Sigma}_{k,k | k-1}^{-1}\\
\label{eq:estep:backwardN}\bs{\Sigma}_{k, k-1 | N_t} &= \bs{\Sigma}_{k, k-1 | k} + \left[\bs{\Sigma}_{k,k | N_t} - \bs{\Sigma}_{k,k | k}\right]\bs{\Sigma}_{k,k | k}^{-1}\bs{\Sigma}_{k, k-1 | k}.
\end{align}

The steps of the expectation-maximization algorithm in terms of the formulae derived in this section now take the following form:
\begin{enumerate}[leftmargin=3ex]
	\item Make an initial guess of $\transition,\,\processcovar,$ and $\torques$.
	\item Find the expectation value of the states using (\ref{eq:estep:forward1})--(\ref{eq:estep:sig_k_k-1}) and (\ref{eq:estep:backward1})--(\ref{eq:estep:backwardN}) using estimates of $\transition,\,\processcovar,$ and $\torques$, either from (i) or a previous iteration of (iii).
	\item Use the state estimates from (ii) to update $\transition,\,\processcovar,$ and $\torques$, using (\ref{eq:mlef})--(\ref{eq:mleq}).
	\item Repeat (ii) and (iii) until log-likelihood converges.
\end{enumerate}
\section{Timing noise amplitudes}
\label{sec:timing_noise_comparison}
In this appendix we compare the white noise torque amplitudes, $\sigma_{\rm c}/\Ic$ and $\sigma_{\rm s}/\Is$, with a statistic commonly used in the literature.

\cite{10.1093/mnras/sty3213} quantified timing noise amplitude in terms of the statistic $S_r$, defined by
\begin{align}
    S_r  = \frac{\langle \sigma_R^2(m, T)\rangle}{T^{2r-1} \langle\sigma_R^2(m, 1)\rangle},
\end{align}
where $\sigma_R^2(m, T)$ is the variance of the residuals left over after subtracting off a polynomial fit of degree $m$, $T$ is the time over which the fit is performed, $r$ is the order of the red noise, and $\langle\sigma_R^2(m, 1)\rangle$ is a correction factor defined in~\cite{1984ApJ...281..482D}. In the latter paper $m-1$ is the degree of polynomial subtracted. $S_r$ is calculated on different time-scales and then a geometric average is performed.

\begin{figure}
    \centering
    \includegraphics[width=0.45\textwidth]{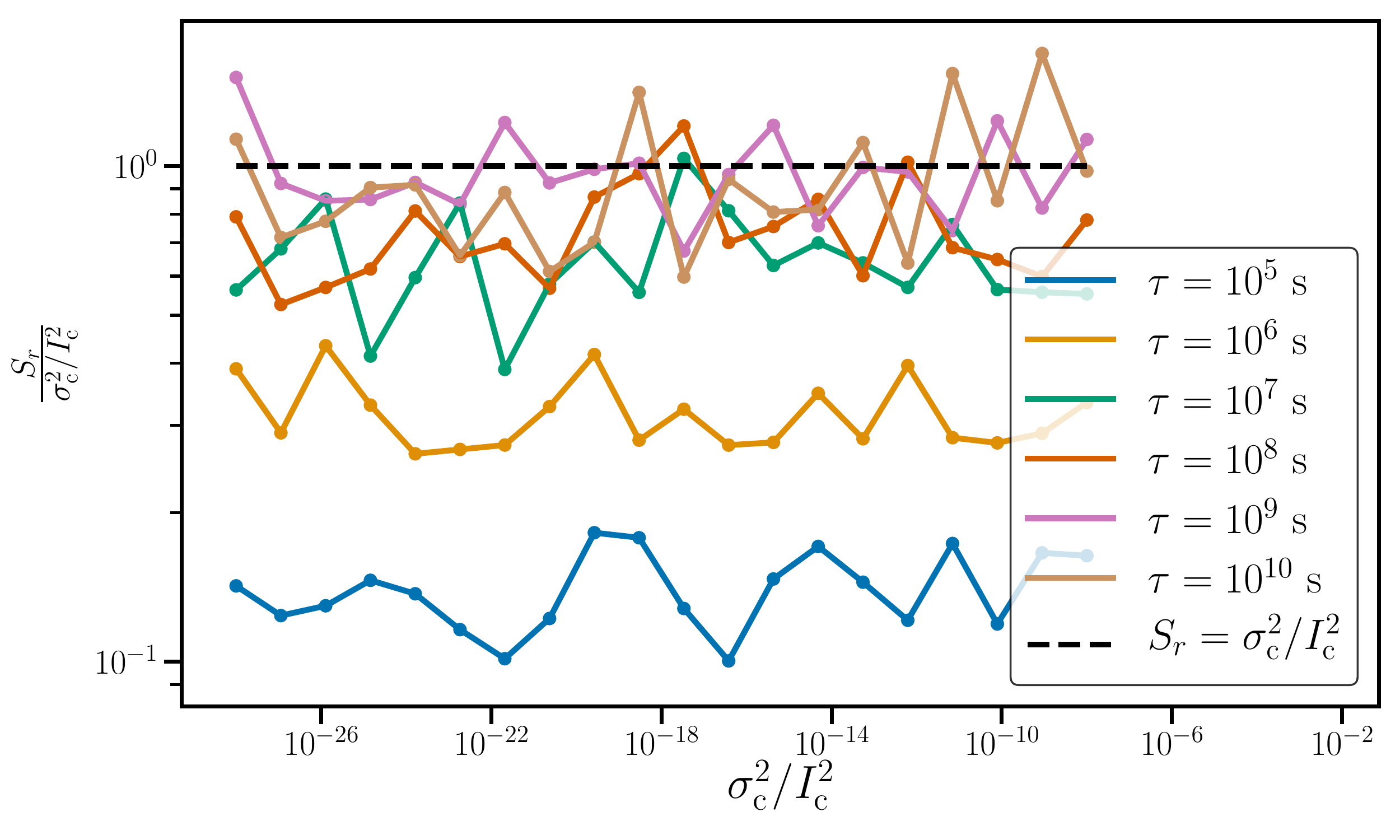}
    \caption{A comparison between the calculated timing noise statistic $S_r$ (defined in the text), and the injected noise amplitude, $\sigma_{\rm c}^2/\Ic^2$ for a range of relaxation time-scales. These statistics are comparable to one another when the time-scale of the relaxation process is long compared to the time between observations.}
    \label{fig:app:noise_study}
\end{figure}

We perform simulations where we numerically integrate~(\ref{EQ:CRUST_EQ_OF_MOTION}) and (\ref{EQ:SUPERFLUID_EQ_OF_MOTION}), subtract off a linear spin-down ($m=1$) and then calculate $S_r$ on twenty logarithmically spaced time-scales, which we then average together geometrically. Performing the polynomial subtraction on the crust angular velocity, $\omgc(t)$, means that we set $r=1$. We do this for a range of $\sigma_{\rm c}/\Ic$ and $\tau$, with $\sigma_{\rm s}/\Is=0$ for simplicity. We show the results of these simulations in Fig.~\ref{fig:app:noise_study}.

In Fig.~\ref{fig:app:noise_study} we plot the ratio between $S_r$ and $\sigma_{\rm c}^2/\Ic^2$. When $\tau$ is large, meaning it takes longer for fluctuations to be damped, $S_r$ and $\sigma_{\rm c}^2/\Ic^2$ are comparable to one another. However, when $\tau$ becomes smaller, and there is noticeable damping, $S_r$ underestimates the amplitude of the intrinsic stochastic process. Plots like Fig.~\ref{fig:app:noise_study} can also be used as a useful heuristic for comparing how we characterize the timing noise in the simulations performed in Section~\ref{sec:simulated_data_results}, with how timing noise is characterized in the literature. In general, $\sigma_{\rm c}^2/\Ic^2$ should be thought of as an upper bound for $S_r$.
\section{Alternative set of injection parameters}
\label{app:alternate_injection_results}
\begin{figure}
    \centering
    \includegraphics[width=0.4\textwidth]{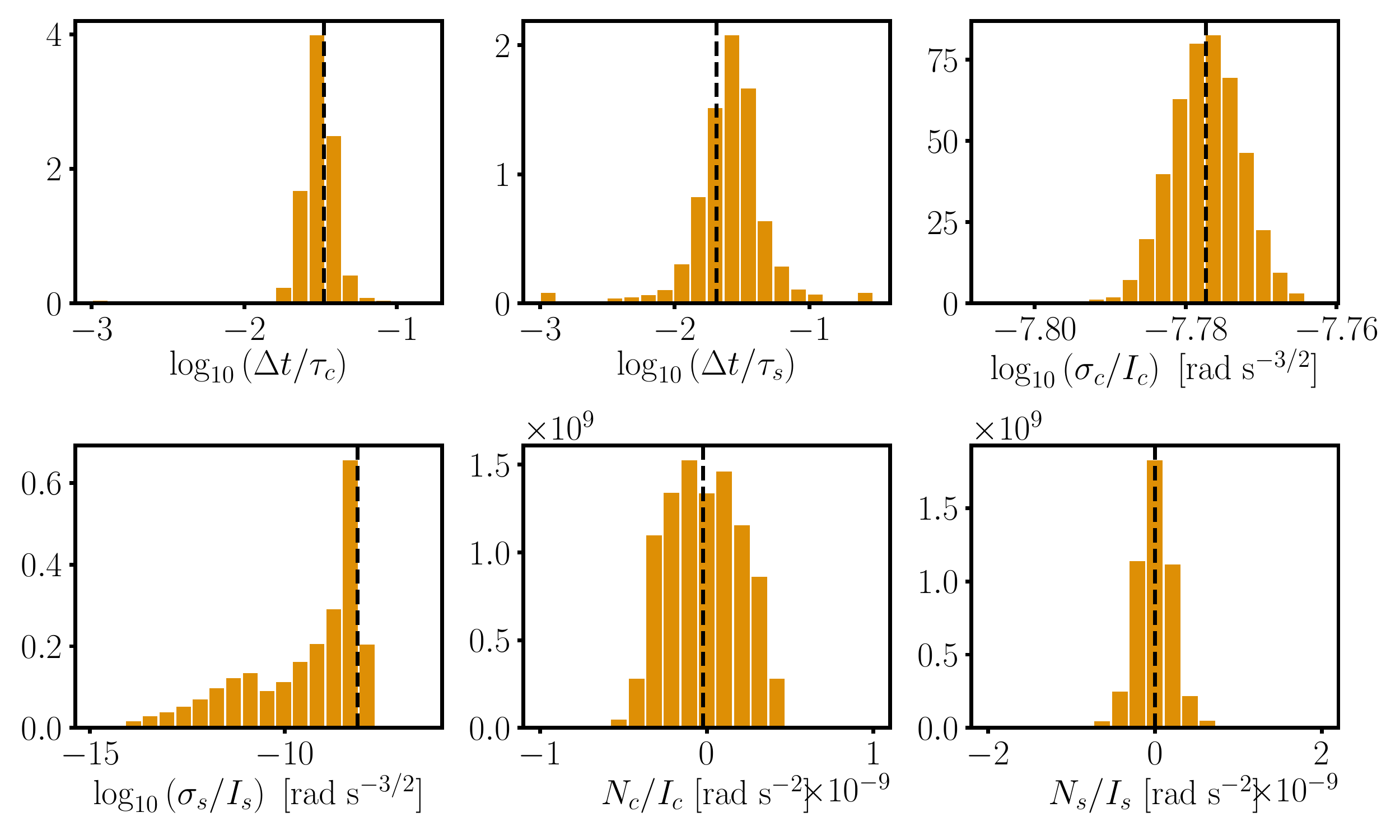}
    \caption{The distribution of estimated parameters for 5000 simulations using electromagnetic only data and $N_t=4630$. The injected parameters are shown in Table~\ref{tab:alternate_injected_parameters} as an alternative to those in the body of the paper. The dashed black lines indicate the injected values.}
    \label{fig:alternate_injection_results}
\end{figure}

In Section~\ref{sec:simulated_data_results} we show results for the set of parameters in Table~\ref{tab:input_params}, which are chosen to mimic an accreting pulsar, hence $\Nc/\Ic >0$. In this section, we choose a different set parameters with $\Nc/\Ic <0$ and $\Ns/\Is<0$, which are more consistent with an isolated pulsar. The set of injected parameters are shown in Table~\ref{tab:alternate_injected_parameters}. For the case where electromagnetic and gravitational-wave data are available, we correctly estimate all of the parameters, as in Section~\ref{sec:simulated_data_results}. For the electromagnetic-only case, the estimated parameters using $N_t=1157$ show similar qualitative results to the example in Section~\ref{sec:simulated_data_results}. There is a peak in $\Delta t/\tauc$ near $10^{-3}$, as well as a peak near the injected value. As with the example in Section~\ref{sec:simulated_data_results}, the peak near $10^{-3}$ disappears for $N_t=4630$. The resulting distribution of estimated parameters for 5000 simulations are shown in Fig.~\ref{fig:alternate_injection_results}. As in the earlier example, we find accurate peaks around $\tauc/\Ic$ and $\sigma_{\rm c}/\Ic$. A peak also emerges around the correct value for $\sigma_{\rm s}/\Is$. However we are still unable to resolve the torques on the individual components.

\begin{table}
{\renewcommand{\arraystretch}{1.2}
\begin{center}
\begin{tabular}{l l l}
\hline
    {\bf Parameter} & {\bf Value} & {\bf Units}\\
    \hline\hline
    $\Nc/\Ic$ & $-2.34\times 10^{-11}$ & $\rm{rad~s^{-2}}$ \\
    $\Ns/\Is$ & $-1.91\times 10^{-13}$ & $\rm{rad~s^{-2}}$ \\
    $\tauc$ & $2.61 \times 10^{6}$ & s \\
    $\taus$ & $4.21 \times 10^{6}$ & s \\
    $\sigma_c / I_c$ & $1.67 \times 10^{-8}$ & $\rm{rad~s^{-3/2}}$\\
    $\sigma_s / I_s$ & $7.36 \times 10^{-9}$ & $\rm{rad~s^{-3/2}}$\\
    $\Omega_c(0)$ & 100 & $\rm{rad~s^{-1}}$\\
    $\Omega_s(0)$ & 100.00001 & $\rm{rad~s^{-1}}$\\
    $t_{k+1} - t_k$ & $86400$ & s \\
    \hline
\end{tabular}
\caption{Parameters used to create an alternative set of simulated data discussed in Section~\ref{app:alternate_injection_results} and shown in Fig.~\ref{fig:alternate_injection_results}. $\omgs(0)$ is chosen such that $\omgc(0) - \omgs(0)$ takes the expected value in~(\ref{eq:simple_model:lag_limit_long_time}). The choice of $\Nc/\Ic<0$ and $\Ns/\Is<0$ is made to simulate a system that is isolated and spinning down.}
\label{tab:alternate_injected_parameters}
\end{center}
}
\end{table}

% Don't change these lines
\bsp	% typesetting comment
\label{lastpage}
\end{document}